\documentclass{aa}  

\usepackage{natbib,graphicx,amssymb,amsmath}
\bibpunct{(}{)}{;}{a}{}{,} 
\newcommand{\kms}{$\rm km\,s^{-1}$} 
\newcommand{\mc}{\multicolumn} 
\newcommand{\tfm}{\tablefootmark}
\newcommand{\tft}{\tablefoottext}

\usepackage{rotating}
\usepackage{sidecap}

\begin{document}
\title{Surviving the hole I:\\ Spatially resolved chemistry around Sgr\,A$^*$}

\author{
  S. Mart\'in\inst{\ref{inst1},\ref{inst2}}
  \and J. Mart\'in-Pintado\inst{\ref{inst3}}
  \and M. Montero-Casta\~no\inst{\ref{inst4}}
  \and P.T. P. Ho\inst{\ref{inst2},\ref{inst5}}
  \and R. Blundell\inst{\ref{inst2}}
}
\institute{
  European Southern Observatory, Alonso de C\'ordova 3107, Vitacura, Casilla 19001, Santiago 19, Chile\\
  \email{smartin@eso.org}\label{inst1}
  \and
  Harvard-Smithsonian Center for Astrophysics, 60 Garden St.,  02138, Cambridge, MA, USA\label{inst2}
  \and
  Centro de Astrobiolog\'ia (CSIC-INTA), Ctra. de Torrej\'on Ajalvir, km. 4, E-28850 Torrej\'on de Ardoz, Madrid, Spain\label{inst3}
  \and
  Department of Astronomy, Columbia University, 550 West 120th Street, New York, NY 10027, USA\label{inst4}
  \and
  Institute of Astronomy and Astrophysics, Academia Sinica, P.O. Box 23141, Taipei 106, Taiwan\label{inst5}
}

\abstract
{
The interstellar region within the few central parsecs around the super-massive black hole, Sgr~A$^*$ at the very Galactic center is composed by a number of overlapping molecular structures
which are subject to one of the most hostile physical environments in the Galaxy.
}
{
Through the study of the morphology and kinematics of the emission from different molecular species as well as the variation of their line ratios we can get key insights on the distribution and interaction
between different gas structures and the energetic phenomena taking place in the surroundings of Sgr A$^*$ as well as the physical processes responsible for the heating in this region.
}
{
We present high resolution ($4''\times3''\sim0.16\times0.11$\,pc) 
interferometric observations of CN, $^{13}$CN, H$_2$CO, SiO, c-C$_3$H$_2$ and HC$_3$N emission at 1.3~mm towards the central $\sim4$\,pc of the Galactic center region.
}
{
{Strong differences are observed in the distribution of the different molecules. The UV resistant species CN, the only species tracing all previously identified circumnuclear disk (CND) structures,
is mostly concentrated in optically thick clumps in the rotating filaments around Sgr~A$^*$.
H$_2$CO emission traces a shell-like structure that we interpret as the expansion of Sgr~A East against the 50~\kms~ and 20~\kms~ giant molecular clouds (GMCs).
We derive isotopic ratios $^{12}$C/$^{13}$C$\sim15-45$ across most of the CND region, but for the northeast arm, 
where the peak of H$_2$CO is observed and ratios $<10$ are found.
The densest molecular material, traced by SiO and HC$_3$N, is located in the southern CND, likely due to shocked gas infalling from the 20~\kms\, GMC streamers, 
and the northeast arm, as a result of the expansion of Sgr\,A~East 
or a connecting point of the 50~\kms streamer east to the CND.
The observed $\rm c-C_3H_2/HC_3N$ ratio observed in the region is more than an order of magnitude lower than in Galactic PDRs.
Toward the central region only CN was detected in absorption.
Apart from the known narrow line-of-sight absorptions, a $90$~\kms\, wide optically thick spectral feature is observed. We find evidences of an even
wider ($>100$~\kms) absorption feature.
Around $70-75\%$ of the gas mass, concentrated in just the $27\%$ densest molecular clumps, is associated with rotating structures 
and show evidences of association with each of the arcs of ionized gas in the mini-spiral structure.}
}
{
These observations provide a coupled
chemical and kinematical picture of the very central region around Sgr~A$^*$. 
Chemical differentiation has been proven to be a powerful tool to disentangle the many overlapping molecular components in this crowded and heavily obscured region.
}

\maketitle

\section{Introduction}
At a distance of $\sim8.0$\,kpc \citep{Reid1993,Groenewegen2008,Ghez2008} the few central parsecs around the super-massive
black hole at the dynamic center of the Galaxy \citep[$4.1\pm0.6\times10^6\,M_\odot$,][]{Ghez2008} is likely one of the
most hostile environments for the interstellar medium (ISM) in the Milky Way.
In this region the molecular gas undergoes strong disruptive tidal forces 
and pervasive X-ray irradiation \citep[$L_X(20-120 \rm keV)\sim2.6\times10^{35} erg\,s^{-1}$;][]{B'elanger2006}
in the vicinity of the black hole.
Likewise, the ISM is subject to supernova remnants (SNR), HII regions,
and strong photodissociating UV fields from the central cluster
\citep[$L\sim1-3\times10^7L_\odot$ and ionizing photon flux of $2\times10^{50}\rm s^{-1}$;][]{Lacy1980,Davidson1992}.

The super-massive black hole is associated with the compact strong non thermal source Sgr~A$^*$.
The mini-spiral of ionized gas consisting of three streamers known as Sgr~A West \citep{Lo1983,Ekers1983}
is likely feeding the central source.
Surrounding Sgr~A West, the so called circumnuclear disk (CND) is actually a clumpy ringlike structure rotating counterclockwise
around the dynamical center \citep{Guesten1987,Wright2001,Christopher2005}.
The molecular ring extends from its inner sharp edge at a projected distance of 1.6~pc from Sgr~A$^*$, out to a smoother edge at $\sim2$~pc \citep{Christopher2005}.
However, the CND appears not to be a single symmetric structure forming a coherent ring \citep{Marshall1995}.
It has been suggested that the CND is a warped structure \citep{Guesten1987} or, due to the incompleteness of the ring, formed by distinct rotating structures
\citep{Jackson1993,Wright2001}.
Most of the observed molecular clumps forming the CND, with diameters of $\sim7''$, are tidally stable and in virial equilibrium \citep{Christopher2005}.
Thus, the rotating molecular structures composing the CND appear to be non transient structures
with an estimated lifetime of $\sim 10^7$~yr \citep{Christopher2005,Montero-Castano2009}.

\begin{figure}[!t]
\centering
\includegraphics[width=0.46\textwidth]{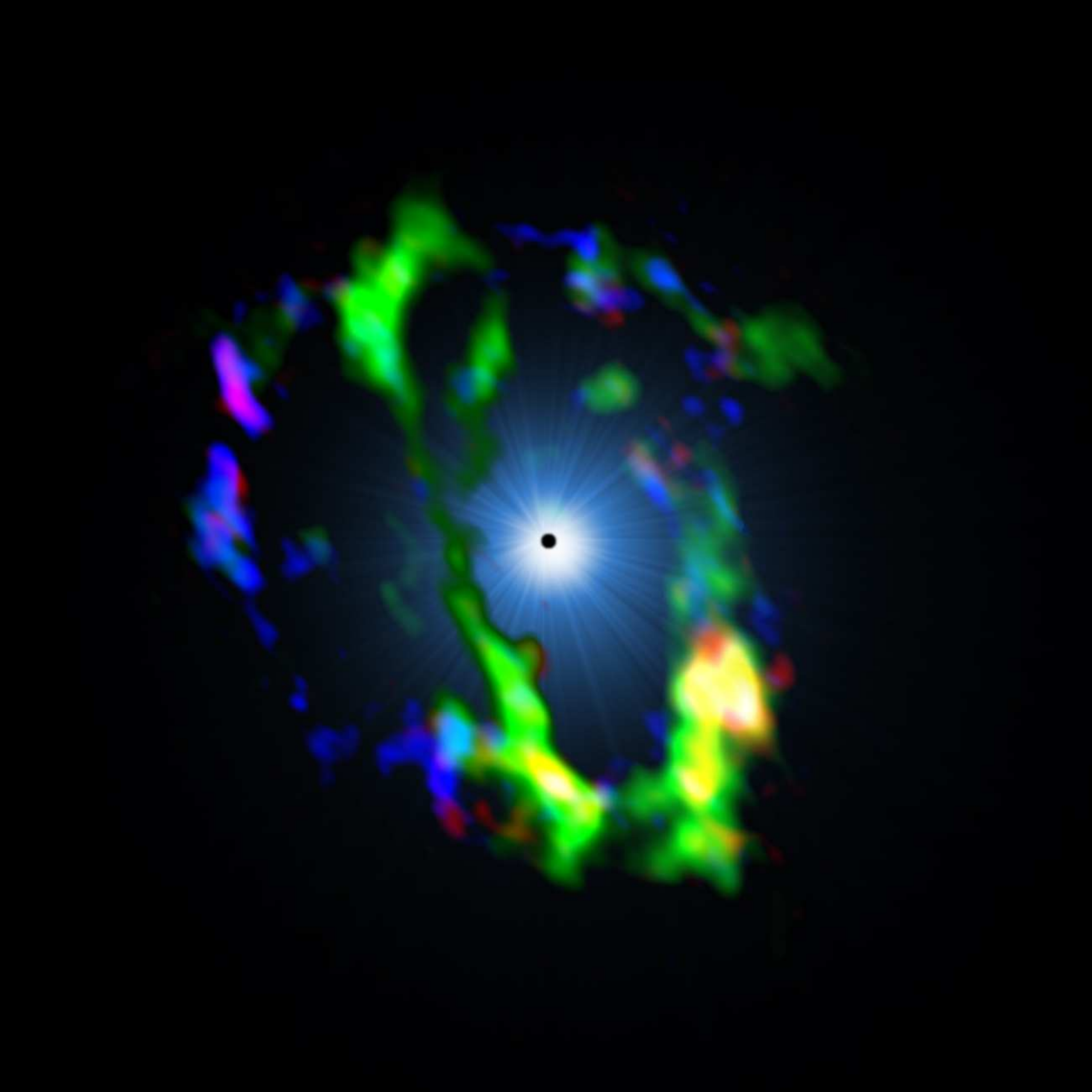}
\caption{Three color composite image on the CND around Sgr~A$^*$ including the emission of SiO (red), CN (green), H$_2$CO (blue).
The central glow shows an artistic representation of the pervasive X-ray and UV-fields from the black hole and central star cluster.
In the center, the black hole position is illustrated with a size corresponding to the map resolution.
\label{fig.Composite}}
\end{figure}

A number of molecular emission observations have been carried out to study the CND structure, its relation to the Sgr~A West streamers and
the possible interaction with the two giant molecular clouds (GMC), namely the 20~\kms\, and 50~\kms\, GMCs \citep[see][for references]{Ho1991,Marr1993,Montero-Castano2009}.
So far, the only species with high resolution aperture synthesis observations are
the HCN $1-0$ \citep{Guesten1987,Wright2001,Christopher2005}, $3-2$ \citep{Marshall1995}, and $4-3$ \citep{Marshall1995,Montero-Castano2009},
the $1-0$ transitions of H$^{13}$CN and HCO$^+$ \citep{Marr1993,Wright2001,Christopher2005}, the CS $7-6$ \citep{Montero-Castano2009}, and
NH$_3$ (1,1), (2,2), (3,3), (6,6) \citep{McGary2001,Herrnstein2005}.
The emission from the high density tracers like HCN, HCO$^+$ and CN nicely trace all the molecular features of the CND, though significant differences
are observed in the line ratios.
These differences are attributed to both abundance and density variations
across the CND \citep[e.g. HCO$^+$/HCN,][]{Christopher2005}.
Although also tracing the CND, the emission of NH$_3$ significantly differs from the emission of the other tracers.
The recent work at low resolution by \citet{Amo-Baladr'on2011} studies, among other tracers, the emission of SiO and HNCO in the central 12 pc around Sgr~A$^*$.
These species show strong abundance variations due to the influence of shocks and UV radiation in this region \citep{Mart'in2008}.
They probe the potential of HNCO/SiO abundance ratio as a measurement of the distance to the central cluster, which provides an insight on the 3-D arrangement
of the different molecular components.

In this paper we report the high resolution ($4''\times3''\sim0.16\times0.11$\,pc) observations of CN, $^{13}$CN, H$_2$CO, SiO, c-C$_3$H$_2$ and HC$_3$N.
As illustrated by the three color composite image shown in Figure~\ref{fig.Composite}, each molecular species traces different molecular components
as a result of the varying physical and chemical conditions in the vicinity of Sgr~A$^*$.

%

\section{Observations}
\label{sect.Observations}
Observations were carried out with the Submillimeter Array
\footnote{The Submillimeter Array is a joint project between the Smithsonian Astrophysical Observatory and the Academia Sinica Institute of Astronomy and Astrophysics and is funded by the Smithsonian Institution and the Academia Sinica.}
\citep[SMA,][]{Ho2004} in Mauna Kea, Hawaii.
We tuned the 230~GHz receivers to observe 2 spectral windows centered at the frequencies
of 217.41~GHz (LSB) and 227.38~GHz (USB).
Correlator configuration provided $2\times1.968$~GHz wide bands at a 0.8125\,MHz spectral resolution.

\begin{table}
\begin{center}
\caption{Observational details of each frequency setup\label{tab.obsdetails}}
\begin{tabular}{c @{\,\,\,} c @{\,\,\,} c @{\,\,\,} c @{\,\,\,} c @{\,\,\,} c}
\hline
\hline
Date              & Config. \tfm{a}  & $\tau_{225}$  \tfm{b} &             \mc{3}{c}{Calibrators}             \\
                  & (\# ant.)        &                       &   Flux        &    Gain            &  Bandpass \\
\hline
2007 Jul 10       &  C (8)           & 0.18                  &   Uranus      &    Callisto        &  3c454.3  \\
                  &                  &                       &               &    J1924-292       &           \\
2007 Jul 15       &  C (8)           & 0.16                  &   Uranus      &    Callisto        &  3c454.3  \\
                  &                  &                       &               &    J1924-292       &           \\
2008 Sep 03       & SC (7)           & 0.25                  &   Callisto    &    J1733-130       &  3c454.3  \\
                  &                  &                       &               &    J1924-292       &           \\
\hline
\end{tabular}
\tablefoot{
\tft{a}{Array configuration (SC: Subcompact - C: Compact) and number of antennae available.}
\tft{b}{Average zenith opacity at 225\,GHz during observations.}
}
\end{center}
\end{table}

A total of 3 SMA observing tracks were acquired in both its compact and subcompact configurations.
Specific observing and calibration details on each of the three SMA tracks
are summarized in Table~\ref{tab.obsdetails}.
Weather conditions were good (PWV$\sim3$\,mm) during compact observations and only mediocre (PWV$\sim4-5$\,mm) during the subcompact.
However, phase was observed to be stable in all three tracks.

The observations in the subcompact configuration data were key to recover the extended emission from the CND structure.
Such flux recovery is perfectly illustrated by comparing the observations of HCN $4-3$ with just the compact configuration data reported by \citet{Montero-Castano2006}
and the final data set including the subcompact configuration data \citep{Montero-Castano2009}.
The combination of both configurations resulted in projected baselines between 8~m (6~k$\lambda$) and 135~m (90~k$\lambda$),
which is equivalent to structure sizes between $34''$ and $2''$, respectively.
Nevertheless, the final cleaned images show negative flux areas nearby the brightest emission regions which are indicative of the significant missing extended ($>30''$) emission.

\begin{figure}[!t]
\centering
\includegraphics[angle=-90,width=0.46\textwidth]{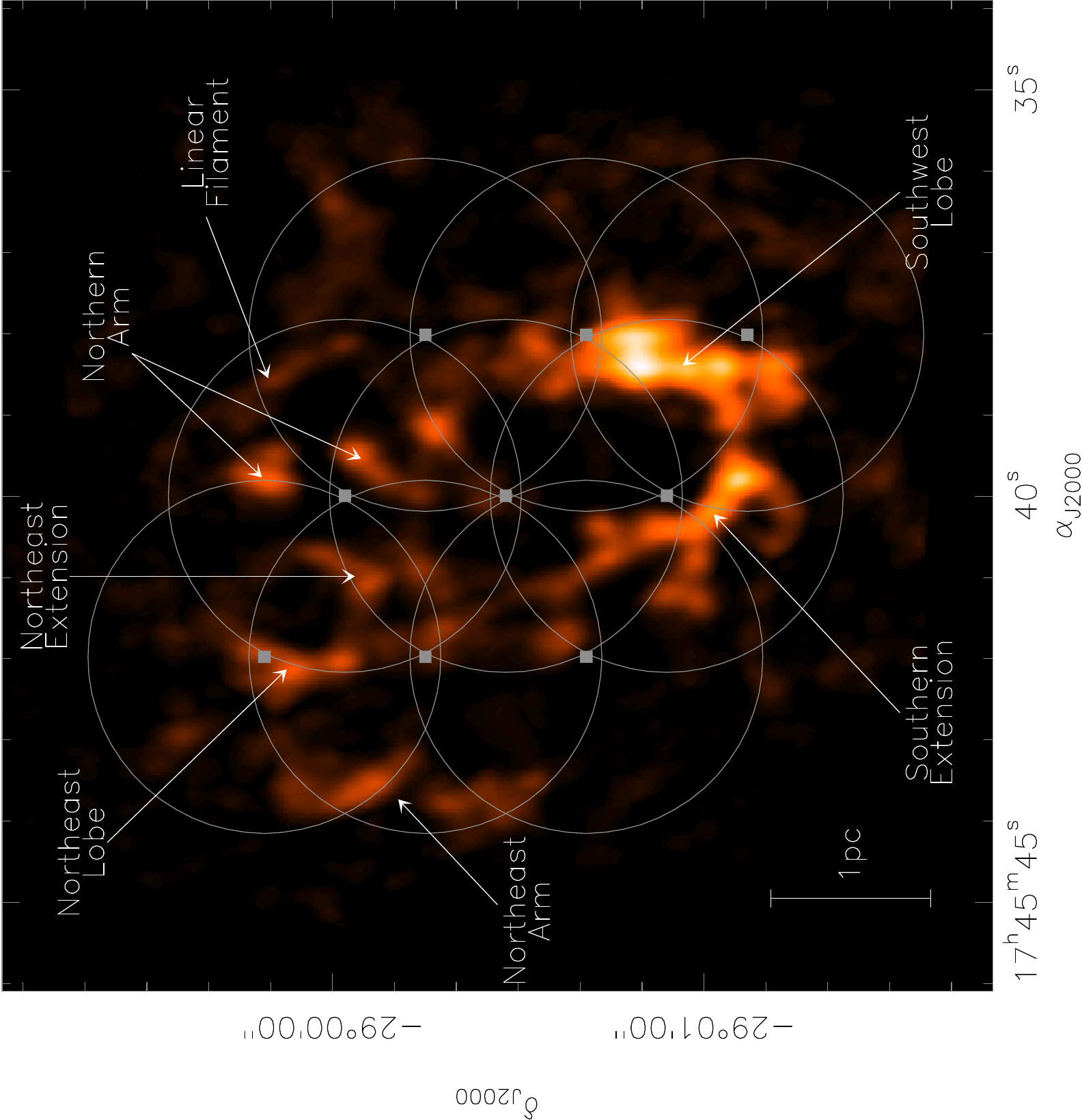}
\caption{Central position and primary beam of each of the 9 field mosaic observations presented in this paper.
Background image is the HCN $4-3$ emission by \citet{Montero-Castano2009}.
The different molecular structures labels are adopted from \citet{Christopher2005} and \citet{Montero-Castano2009}.
\label{fig.mosaic}}
\end{figure}

We observed a 9 pointing mosaic centered at $\alpha_{\rm J2000}=17^{\rm h}45^{\rm m}40^{\rm s},\,\delta_{\rm J2000}=-29^\circ00'28''$
and arranged with a 26$''$ spacing and a declination offset between each 3 point column of $13''$ as shown in Fig.~\ref{fig.mosaic}.
This mosaic configuration was aimed to fully sample the elongated CND structure and
all the features identified in the high resolution maps of HCN $4-3$ by \citet{Montero-Castano2009}.
The observation was performed in short 4 minutes integration on each of the mosaic pointings with
3 minutes integration on the gain calibrators every $\sim$20 minutes.
The combination of the 3 tracks resulted in a total integration time per pointing of $\sim84$ minutes.

Continuum subtraction was performed in the UV plane using the available line free channels between all the identified molecular features.
Data calibration and reduction were performed using the MIR-IDL package, while we used MIRIAD~\citep{Sault1995} for cleaning and GILDAS \footnote{http://www.iram.fr/IRAMFR/GILDAS}
for data display and analysis.
The natural weighting of the visibilities resulted in a beam size of $4.1''\times2.9''$ with a position angle of $\sim30^\circ$.
The theoretical rms noise for individual pointings (assuming no overlapping) was $\sim40\rm mJy\,beam^{-1}$ in 10~\kms\, channels.
The final achieved rms noise was $\sim 31$ and $34\rm\,mJy\,beam^{-1}$ for the LSB and USB, respectively, as measured in the central $40''$ region in line free channels. 

\begin{table}
\begin{center}
\caption{Spectral parameters of the transitions detected \label{tab.molectrans}}
\begin{tabular}{c c c c c}
\hline
\hline
Molecule         & Transition          &  Frequency        & $E_{upper}$ \tfm{a}   &  $A_{ul}$  \tfm{b}        \\
                 &                     &    (GHz)          &   (K)                 &  ($10^{-4}$\,s$^{-1}$)    \\
\hline
SiO              & $5-4$               &  217.104          &   31                  &  5.2                      \\
$^{13}$CN \tfm{c}& $2-1$               &  217.467          &   16                  &  1.0                      \\
c-C$_3$H$_2$     & $6_{1,6}-5_{0,5}$   &  217.822          &   39                  &  5.9                      \\
H$_2$CO          & $3_{0,3}-2_{0,2}$   &  218.222          &   21                  &  2.8                      \\
CN \tfm{c}       & $2-1$               &  226.875          &   16                  &  1.1                      \\
HC$_3$N          & $25-24$             &  227.419          &   142                 &  9.3                      \\
\hline                  
\end{tabular}           
\tablefoot{             
\tft{a}{Energy of the transition upper level.}
\tft{b}{Einstein coefficient}
\tft{c}{Frequency and $A_{ul}$ refer to the brightest transition of the hyperfine group.}
}
\end{center}
\end{table}

\begin{figure}[!t]
\centering
\includegraphics[angle=-90,width=0.46\textwidth]{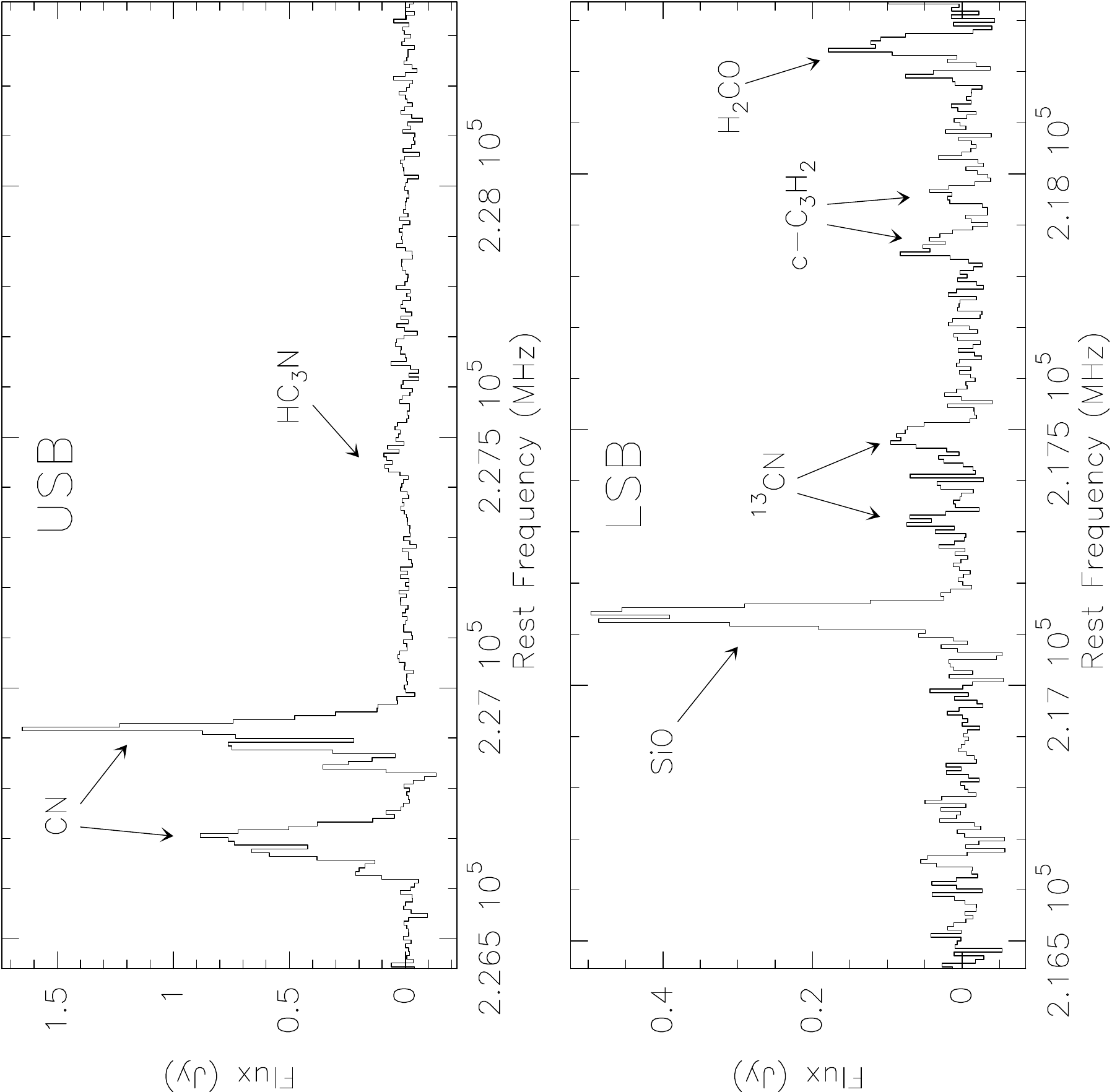}
\caption{Sample spectra extracted from a $4''$ size square region at the position 1 indicated in Fig.~\ref{fig.Integrated}.
The spectra of both USB and LSB are shown with the identified spectral transitions labeled.
\label{fig.spectraSample}}
\end{figure}

\section{Results}
\label{sect.results}

\begin{figure*}[!t]
\centering
\includegraphics[angle=-90,width=\textwidth]{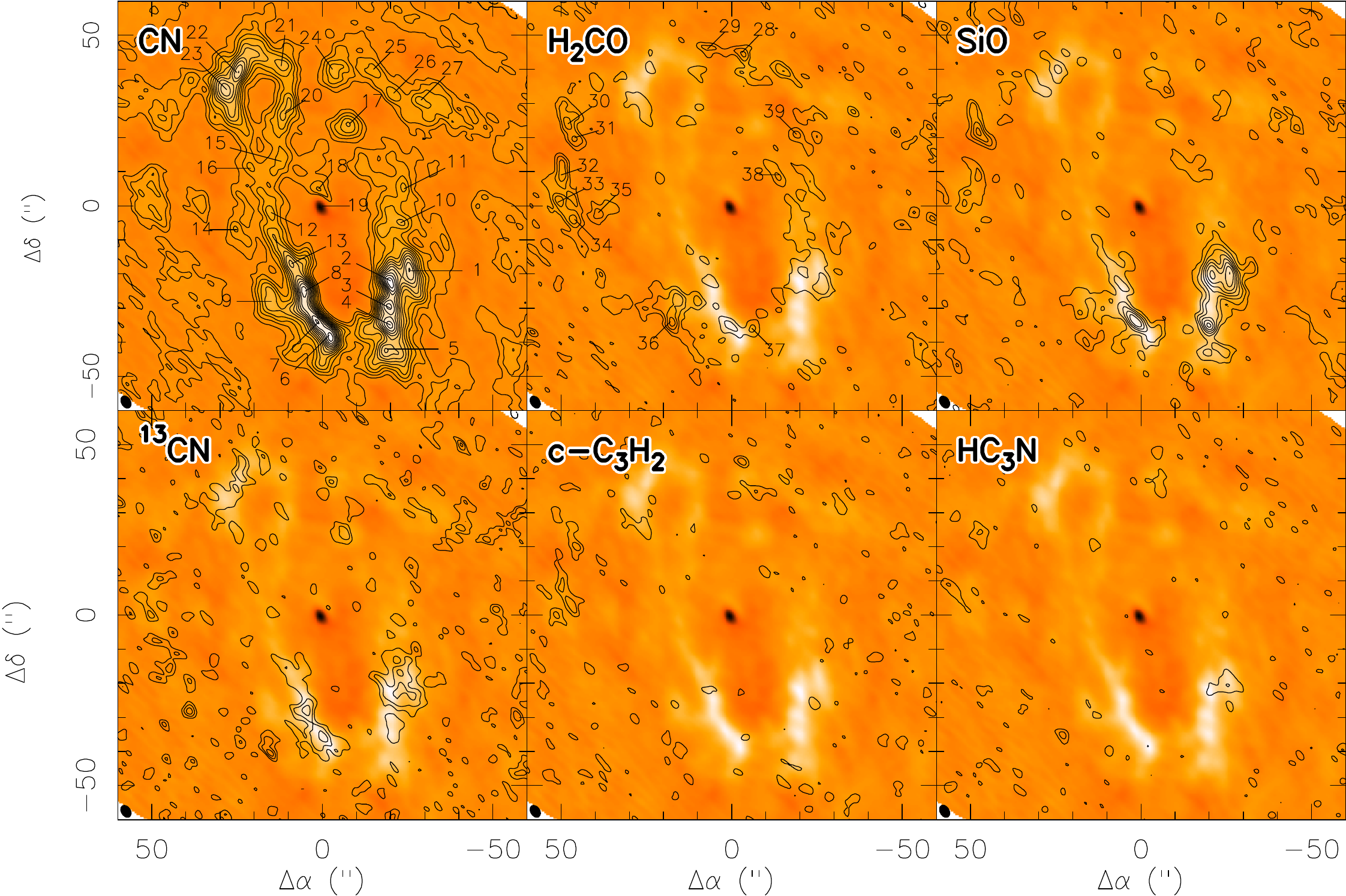}
\caption{Natural weighted integrated intensity maps of all detected species in this work.
Beam size of $4.1''\times2.9''$ is shown in the lower left corner of each box.
Emission was integrated over the velocity range between $-140$ and $+140$ \kms for SiO, C$_3$H$_2$, and  HC$_3$N, while we used the velocity ranges of
[$-190,+470$] for CN, [$-80,+140$] for H$_2$CO, and [$-190,+350$] for $^{13}$CN
(See Sect.~\ref{sect.results} for details).
Background image corresponds to the integrated emission of CN.
Contour levels are at $2 \sigma$ (with $\sigma=2.3\rm\,Jy\,beam^{-1}$\,\kms) steps for all maps
but for CN, where the first contour is $3 \sigma$ and increases in $8 \sigma$ steps.
\label{fig.Integrated}}
\end{figure*}

Within the total $\sim4$~GHz frequency band observed, we have detected the emission of 6 molecular species.
Fig.~\ref{fig.spectraSample} shows the sample spectra extracted from the resulting data cubes at the position of the brightest molecular emission 
(position 1 in Fig.~\ref{fig.Integrated}) where all of the identified spectral features are labeled.
Basic spectroscopic details for each molecular transition are given in Table~\ref{tab.molectrans}.
Additionally, the emission of HC$_3$N $J=24-23$ emission at 218.3~GHz is detected close to the H$_2$CO transition. However, this line
is truncated at the lower velocities and therefore has not been included in this work.

In Fig.~\ref{fig.Integrated} we present the integrated intensity maps from each of the detected transitions.
The emission of SiO, C$_3$H$_2$, and  HC$_3$N was integrated in the velocity range between $-$140~\kms\, and $+$140~\kms\,, where emission is detected.
In the case of CN, we have integrated between $-$190~\kms\, and $+$470~\kms\, so that the two groups of transitions are included in the image.
Similarly, $^{13}$CN emission was integrated from $-$190~\kms\, up to $+$350~\kms, where SiO starts showing emission.
Finally, H$_2$CO was integrated between $-$80~\kms\, and $+$140~\kms\, to avoid the significant emission from HC$_3$N $24-23$  (@ 218.324 GHz)
at the lowest velocities.
The narrow features integrated over a relatively wide velocity range caused the integrated maps of the faint $\rm c-C_3H_2$ and HC$_3$N to be mostly dominated by noise.
Thus, we have included channel maps of integrated emission over 30~\kms~ bins for each species in Figs.~\ref{fig.channelCN} to~\ref{fig.channelHC3N} (online) where the origin of
the fainter emission is clearly identified.

In Fig~\ref{fig.Integrated} we have labeled a number of selected positions corresponding to the molecular cores identified in the integrated maps.
We have extracted the spectra from each of these positions and performed a multiple velocity component Gaussian fit.
The resulting parameters from these fits are shown in Table~\ref{tab.CNvs13CN} (online) for CN and $^{13}$CN and
in Table~\ref{tab.GaussFits} (online) for the rest of species.


\section{Spatially Resolved Chemistry}
The differences in the distribution of the emission of each of the detected species are evidenced in the integrated maps in Fig.~\ref{fig.Integrated}.
Strong differences are also evident in velocity as shown in the channel maps in Fig.~\ref{fig.channelCN}$-$\ref{fig.channelHC3N} (online).
Below we describe the main morphological structures observed for each individual species.
The origin of each species is discussed in terms of chemical differentiation and excitation.
In order to make a detailed study of the chemical differentiation we show line ratio maps in 30~\kms~ velocity bins in Figs~\ref{fig.channelCNvsH2CO}-~\ref{fig.channelSiOvsHC3N} (online).

The labeling of the different CND molecular structures, shown in Fig.~\ref{fig.mosaic}, are adopted from \citet{Christopher2005} and \citet{Montero-Castano2009}.

\subsection{CN: Overall molecular emission}
\label{sec.CN}
\begin{figure}[]
\centering
\includegraphics[angle=-90,width=0.46\textwidth]{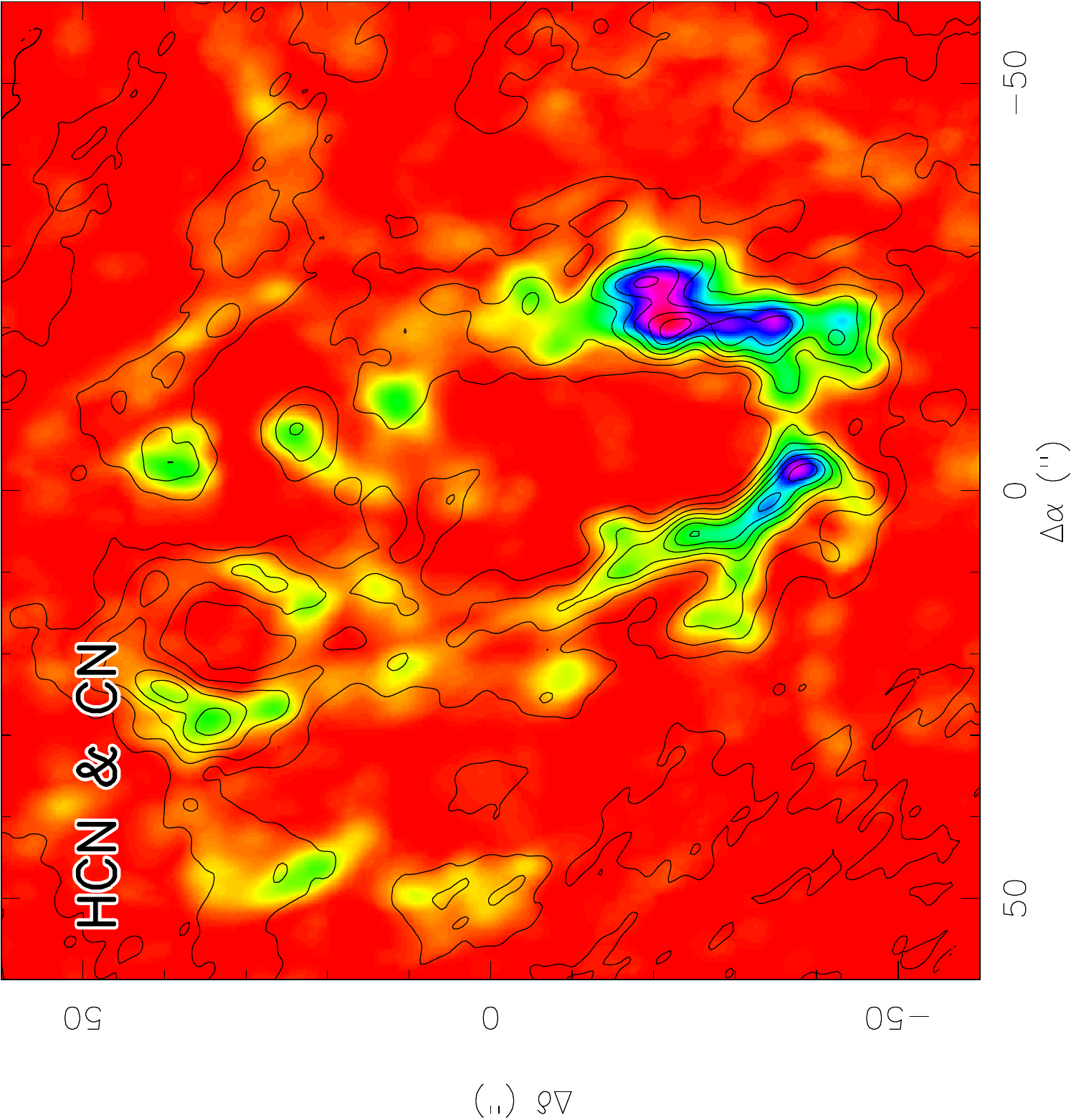}
\caption{Comparison between the HCN $4-3$ integrated emission \citep[color;][]{Montero-Castano2009}
and the CN integrated emission (contours). Contour levels start at $3\sigma$ and increases
in $10\sigma$ steps.
\label{fig.HCNvsCN}}
\end{figure}

Out of the six molecular species observed in this work, CN is the only one tracing all previously identified CND features.
Though differences are observed, the integrated emission of CN appears to be mostly coexistent and tracing the same molecular material
as that observed in HCN and CS \citep{Christopher2005,Montero-Castano2009}.
In Fig.~\ref{fig.HCNvsCN} we present a comparison between the HCN $4-3$ integrated emission from \citet{Montero-Castano2009} and that of CN $2-1$.
Most of the main formation paths of CN in gas phase involve HCN, either by direct photodissociation of HCN into CN, charge exchange with CN$^+$, ion-neutral reactions with HCN$^+$ and $\rm C_2H^+$ or by
dissociative recombination of HCN$^+$ \citep{Woodall2007}.
These processes are expected to be efficient due to the 
enhancement of ion species by the radiation which pervades the environment surrounding the central cluster.
Moreover, as shown in Fig.~\ref{fig.photorate}, CN is the most UV resistant species among those presented in this paper, with photodissociation rates for moderate shielding similar to those
of HCN and CS.
Indeed, CS formation is also known to be favored in UV radiated regions via the enhancement of its main precursor S$^+$ \citep[][and references therein]{Goicoechea2006},
which may explain why the three species show a similar spatial distribution in this region.

We can, however, point out a number of significant differences between the HCN
and CS emission and that of the CN maps.
The emission in both the southwest lobe and southern extensions agree well in extension and location of the emission peaks of CN and HCN.
Even though CN $2-1$ and HCN $1-0$ transitions share similar excitation conditions, the southern extension observed in CN extends towards
the north connecting with the northern extension. This is observed in the $4-3$ transition of HCN
as well as in CS $7-6$ \citep{Montero-Castano2009}, but not in HCN $1-0$ \citep{Christopher2005}.
Thus, this is not an effect of self-absorption of the HCN $1-0$ in this extension towards the north but rather filtering of the extended
emission in the OVRO maps by \citet{Christopher2005}.
The northeast extension and lobe appear connected in the CN map, similar to that observed in HCN $1-0$. Such connection is not observed
in the $4-3$ emission, likely due to a higher excitation in the southern part of the northeast extension as shown by the HCN $4-3/1-0$ ratio
in Fig.~8 from \citet{Montero-Castano2009}.
The CN emission in the northeast arm is significantly less prominent than that observed in HCN which might point out towards a real abundance
deficiency in this elongated feature. As discussed below (Sect.~\ref{sec.H2COshell}), this structure is prominent in H$_2$CO and matches the emission structure
traced by HCN.

Based on the HCN $4-3$ data of \citet{Montero-Castano2009}, at a similar resolution as the CN $2-1$ data presented in this work, we derive some rough estimates of
the HCN/CN ratio of $\sim3-2$ in the southern CND down to a slightly lower ratio of $\sim2-1$ in the northeast lobe.
The ratios have been estimated by comparing the integrated intensity maps and confirmed by the comparison of the peak intensities of the spectra in selected positions shown
in Fig. 4 from \citet{Montero-Castano2009}. 
These difference could be explained by the different excitation conditions along the CND region (Fig.~8 in \citet{Montero-Castano2009}).

\subsection{H$_2$CO: Expanding shell}
\label{sec.H2COshell}
Fig.~\ref{fig.Integrated} shows how the integrated intensity of H$_2$CO differs drastically from that of CN.
Even though H$_2$CO emission is detected towards the positions of peak CN emission (see Table~\ref{tab.GaussFits} online),
most of the emission of this molecule is detected in the outer regions of the CND.
The emission of H$_2$CO appears to trace a shell-like structure roughly centered at an offset position of
$\sim(+12'',+10'')$ from Sgr~A$^*$ with a radius of $\sim37''$.
These differences are particularly evident in the channel maps in Fig~\ref{fig.channelH2CO} (online).
For the velocities below 0~\kms, where significant high velocity emission from HC$_3$N $24-23$ is detected in the northeast arm, the H$_2$CO emission follows the brightest
regions observed in CN.
At these velocities, CN/H$_2$CO line ratios range from  $3-4$ in the southwest lobe and in its connection to the southern extension (position 37 in Fig.~\ref{fig.Integrated}) up to $>30$ in the regions where H$_2$CO is not detected.
However the distribution of the emission differs for positive velocities. Channel maps centered at 0 and 30~\kms~ show emission mostly towards the south, while channel maps
centered at 30 and 60~\kms~ show strong emission in the eastern most region of the maps.
In these velocity channels the H$_2$CO emission is observed to be extended over regions where CN is not detected, where line ratio limits CN/H$_2$CO$<0.2$ are found.
This is particularly prominent for the emission in the 0 and 30\,\kms~centered channels where strong H$_2$CO emission is observed east of the southern extension.

\begin{figure}[]
\centering
\includegraphics[width=0.46\textwidth]{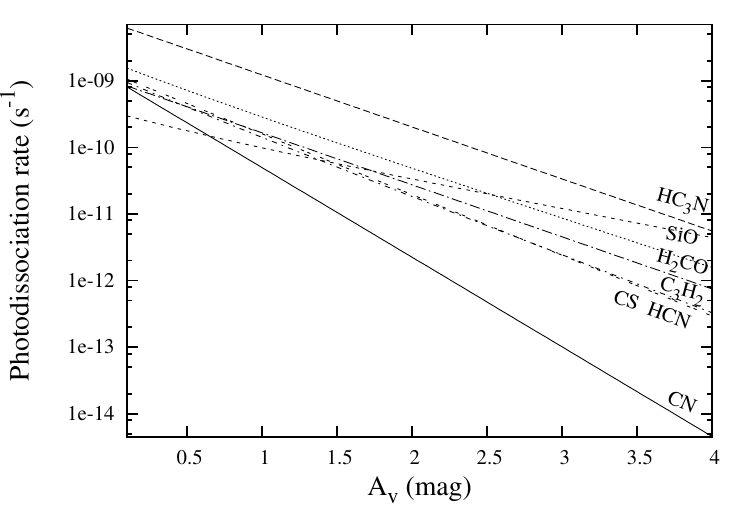}
\caption{Photodissociation rate as a function of the visual extinction for each of the species discussed in this paper.
Rates are extracted from UMIST database \citep{Woodall2007} and assume the standard ISM UV field \citep{Draine1978}.
\label{fig.photorate}}
\end{figure}

The H$_2$CO molecule is efficiently formed in dust grains \citep{Watanabe2002,Watanabe2004,Cuppen2009}
where photo-induced chemistry by UV radiation can lead to the production of H$_2$CO from the H$_2$O and CO in solid phase \citep{Shalabiea1994}.
Thus, the abundance of H$_2$CO in the gas-phase can be enhanced due to grain erosion by shocks and/or thermal evaporation in the vicinity
of hot young stars \citep{Shalabiea1994}, while the main destruction paths are via photodissociation and reactions with C$^+$ \citep{Federman1991,Turner1993}.
Indeed, after HC$_3$N, the H$_2$CO molecule is the most sensitive to photodissociation in our sample, which explains why, apart from the shell-like structure, 
H$_2$CO is only detected closer to Sgr~A$^*$ towards the brightest CN regions. In these dense regions the molecular gas is more shielded from photodissociation.

The observed circular structure could be interpreted as the interaction of Sgr~A East mostly with the 50~\kms\, and 20~\kms\,
giant molecular clouds, east and south of the CND, respectively.
The kinematics of H$_2$CO also support this scenario as seen in the channel maps. Taking the Gaussian fits in Table~\ref{tab.GaussFits} (online),
the brightest observed H$_2$CO molecular components have velocities of $30-60$~\kms\, towards the northern half of the CND with its maximum towards the north east,
and velocities of $\sim10$~\kms\, towards the southern regions. These velocities are consistent with the emission of H$_2$CO being associated with an expanding shell impacting the 50~\kms\, GMC in the north and northeast of the CND,
and the 20~\kms\, GMC in the southern region.
In these regions, H$_2$CO abundance would be enhanced in the gas-phase by the expansion of Sgr~A East into the surrounding GMCs, where it will survive in moderately dense molecular clumps 
\citep[$\rm A_v>2$,][and Fig.~\ref{fig.photorate}]{Shalabiea1994},
shielded from the pervading UV radiation from the central cluster.
The compression by an expanding shell is supported by the enhanced emission of HC$_3$N and SiO in this structure (Sect.~\ref{sec.SiOshocks} and ~\ref{sec.HC3Ndensest}).

Chemical differentiation of the eastern region is also suggested by the lower $^{12}$C/$^{13}$C ratio measured in this region as compared to the central CND structure (see Sect.~\ref{sect.1213C}).

\subsection{SiO: Infalling Shocked Material}
\label{sec.SiOshocks}

With a critical density of $\sim10^8~\rm cm^{-3}$, the $5-4$ transition of SiO is tracing the densest molecular material in the CND.
The single-dish maps of the SiO $2-1$ \citep{Amo-Baladr'on2011} show the SiO emission to be extended and tracing the whole CND structure.
Similarly, the interferometric map of SiO $2-1$ by \citet{Sato2009} clearly shows emission along the CND.
However, our map of the $5-4$ transition emission is observed to be more compact as expected due to the different excitation conditions between both transitions.
Indeed, the one order of magnitude larger critical density of the $5-4$ transition compared to that of the $2-1$ makes it unlikely for the emission to be
spread over large scales (see Sect.~\ref{sect.Observations}) so as to be
filtered out in our maps.

SiO $5-4$ emission is mostly found towards the southern region of the CND, in both the southwest and southern extension, and the northeast arm.
We find CN/SiO line ratios ranging between $1-5$ in the southern region and of $\sim8-10$ in the northeast. Similar to H$_2$CO, lower limits of $>30$ are found in the regions with no SiO detection.
SiO is also observed in the northeast arm and east of the southern extension, where prominent H$_2$CO emission was found, with limits to the CN/SiO$<0.3$ in the regions with no CN detection.

The SiO molecule is not efficiently formed in the cold gas phase if Si is not released to gas phase. Thus, its emission is commonly accepted to be a tracer of the presence of shocks responsible for SiO ejection from grain mantles \citep{Ziurys1989,Martin-Pintado1992}.
Indeed, SiO is widely spread across the Galactic center region likely due to large-scale shocks affecting the giant molecular clouds \citep{Martin-Pintado1997}.
Though SiO has been found to be correlated with the 6.4~keV Fe line, which points out to an X-ray induced enhancement of SiO \citep{Mart'in-Pintado2000,Amo-Baladr'on2009},
the lack of prominent 6.4~keV line emission in the CND region \citep{Yusef-Zadeh2007} and the compact emission of SiO makes it unlikely for the X-ray to be the main driver of the high SiO abundance.
Though UV photodesorption has been proposed as a possible driver of the SiO enhancement \citep{Schilke2001} one would expect its distribution to be more uniform within the central CND, similar to CN and HCN, and not so prominent
in the northeast arm.
Moreover, even if it was claimed that shock velocities $>40$~\kms~ are needed to destroy the silicate grains \citep{Seab1983}, recent observations and modeling have shown low velocity C-shocks ($v=5-10$\kms) to be efficiently sputtering the silicon
from grain mantles \citep{Jim'enez-Serra2005,Jim'enez-Serra2008}.
Thus, the enhancement of SiO towards the southern region could be understood by shock destruction of dust grains in the cloud complexes \citep{Amo-Baladr'on2011}.
From this material, the $5-4$ transition is tracing the densest molecular material affected by shocks.
Such shocks could be originated by gas inflow into the CND.


\begin{figure}[]
\centering
\includegraphics[width=0.45\textwidth]{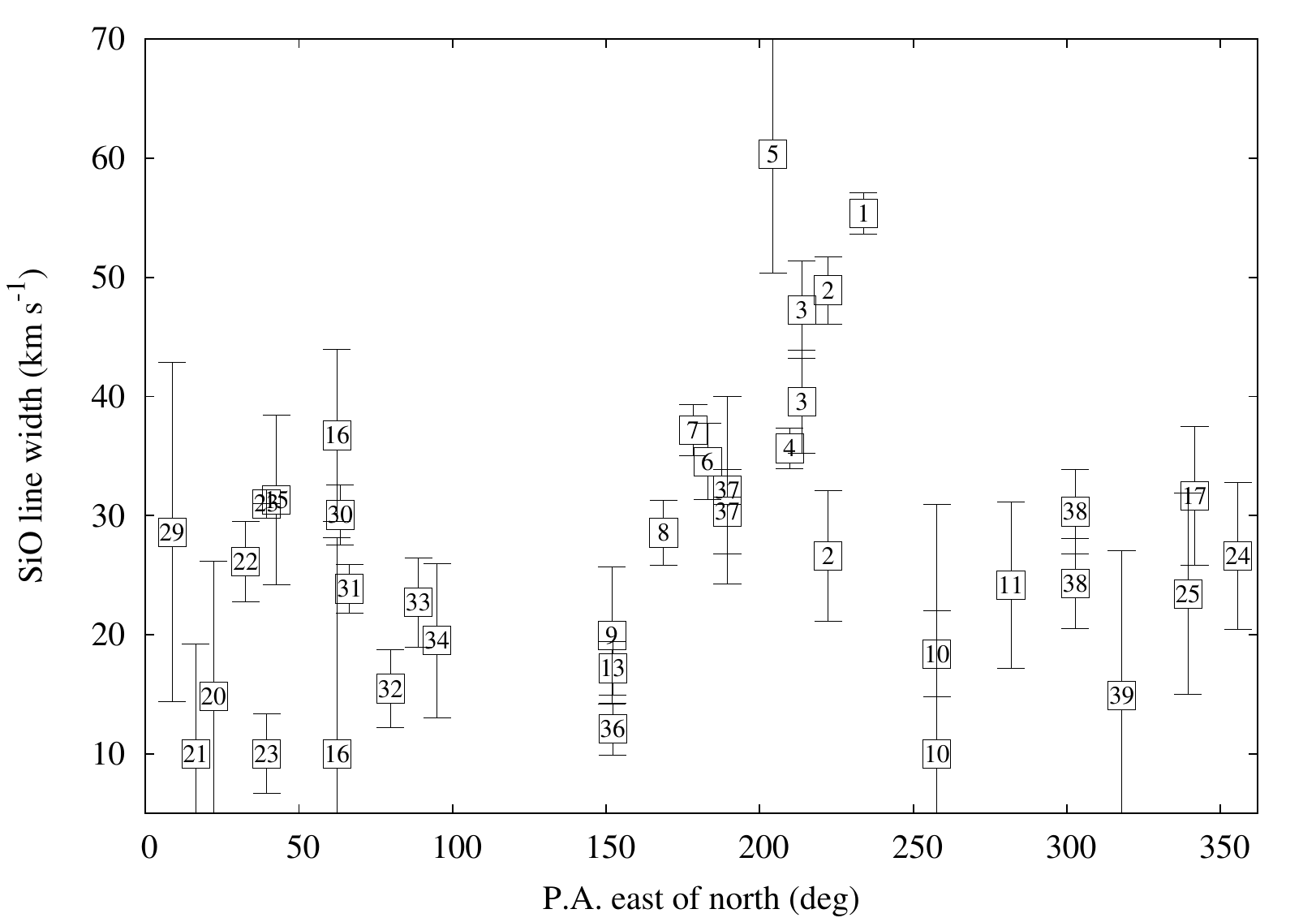}
\caption{SiO line width as a function of the position angle on the sky east of north as measured from Sgr~A$^*$.
Numbers correspond to Fig.~\ref{fig.Integrated} labels.
Position 14 and one of the velocity components in position 9 are not shown due to the large uncertainty in the width measurement.
The positions with the larger widths at P.A. around $200^\circ$ correspond to the southern lobe and southern extension of the CND structure.
\label{fig.SiOwidth}}
\end{figure}

\citet{Coil1999} found kinematic evidences of gas streamers from the southern 20~\kms\, molecular cloud into the CND.
The main gas flow, known as the {\it southern streamer} would be connecting to the southern extension and could be currently feeding this molecular structure.
Moreover, \citet{Coil1999} propose the presence of two streamers emerging from the 20~\kms\, cloud. The second streamer would originate in the northwest of the 20~\kms\, cloud and would connect the CND towards the southwest lobe.
The SiO $2-1$ map by \citet{Sato2009} shows emission from two filaments south of the CND which supports the idea of the two streamers.
The presence of these streamers connecting the 20~\kms\, cloud with the CND would explain the two separated molecular complexes observed in the southern CND as well as the enhancement of SiO emission due
to shocked material in the interaction between the streamers and the southern structures.
This gas infall might be the cause for the wider line widths observed towards the southern lobe. Fig.~\ref{fig.SiOwidth} shows the measured line widths on each of the molecular components studied, where we observed
the line widths to vary between 10 and 30~\kms\, in most positions but for those in the southern regions of the CND where line widths are systematically larger than 30~\kms, up to values of 60~\kms.
This increase in the linewidth is consistent with the one observed in NH$_3$ as the gas approaches the CND from the 20~\kms\,GMC \citep{Coil1999}.
We note how the SiO emission, even though following the kinematics of CN (see Sect.~\ref{sec.kinematics}), is observed to be in the outer edge of the southwest lobe as seen in the $-120$ to $-90$~\kms channel maps (Fig.~\ref{fig.channelCNvsSiO} online),
while the SiO enhanced gas is merged with the gas traced by CN in the $-30$ and 0~\kms channel maps.
Similarly, the SiO emission in the southern extension is observed towards the southern edge of the structure.
Towards the northeast lobe we also observe an enhancement of the SiO emission right at the center of this structure which could be linked to the 50~\kms\, streamer from the nearby GMC \citep{Jackson1993}.
Though SiO enhancement is observed at the positions where the inflowing streamers are claimed to connect to the CND, 
in order to fully confirm the inflowing origin of the SiO emission we need large scale sensitive mapping, such as the one carried out towards the southern part of the CND by \citet{Sato2009}.

Additionally we observe SiO emission in the northeast arm and east of the southern extension following closely the emission observed in H$_2$CO (see Sect.~\ref{sec.H2COshell}).
Silicon sputtering is expected at these positions in the scenario of Sgr~A East expanding against the 50~\kms\, and 20~\kms\, GMCs.

It is important to note that the line widths of SiO at all measured positions do not show significant differences, within the measurement errors, with the line widths of the other species.
Moreover, we do not find traces of shock induced differences in the line profiles, such as line wings. 
Even if the velocity resolution of our maps is not enough to detect asymmetries caused by low velocity C-shocks, 
we can discard high velocity shocks as the main origin of the SiO abundance.
Thus, SiO is likely enhanced by cloud-cloud collisions as the material approaches the CND.


\subsection{$^{13}$CN: Carbon $^{12}$C/$^{13}$C Isotopic Ratio}
\label{sect.1213C}
The simultaneous observation of $^{12}$CN and $^{13}$CN provide a direct measurement of the $^{12}$C/$^{13}$C isotopic ratio in the inner
region of the Galaxy.
Additionally, the fit of the hyperfine molecular components to the observed profiles allows us to estimate the opacity of the CN emission
along the CND.
Table~\ref{tab.CNvs13CN} (online) shows the results of the fit to the CN and $^{13}$CN observed profiles in a number of selected positions labeled in Fig.~\ref{fig.Integrated}.
The opacity of the brightest component of the CN hyperfine group, derived from the observed line profiles, is also tabulated in Table~\ref{tab.CNvs13CN} (online).
We find that CN is optically thick ($\tau\gtrsim 1$) throughout most of the CND structure.
We have calculated the integrated intensity ratio CN/$^{13}$CN in all positions where $^{13}$CN was detected.

From our estimate of the opacity of CN, we calculated the opacity-corrected CN/$^{13}$CN as a proxy to the $^{12}$C/$^{13}$C isotopic ratio.
As shown by comparing the last two columns in Table~\ref{tab.CNvs13CN} (online) and
given that only the brightest hyperfine transitions within the observed CN spectral features are severely affected by opacity,
the integrated intensity ratio CN/$^{13}$CN appears to provide a reasonable good estimate of
the carbon isotopic ratio. 
We measure $^{12}$C/$^{13}$C ratios in the wide range of $15-45$. This result is in agreement with the values reported towards the Sgr~A region
\citep{Audouze1975,Wannier1980} and consistent with the scenario of $^{13}$C enrichment towards the Galactic center region \citep{Wilson1994,Wilson1999}.

Only a few positions are observed to have significantly low isotopic ratios $<10$. However these positions are, either particular velocity components which might not be
related to the CND structure, or the eastern-most molecular component in our maps where the H$_2$CO is most prominent.
This difference in the isotopic ratio seems to support the H$_2$CO molecular complex to be not only morphological (Sect.~\ref{sec.H2COshell}) 
but chemically detached from the central CND structure.
However, the number of positions with measurements of this ratio are too limited to be able to draw firm conclusions based on the different
carbon isotopic ratios.

\subsection{c-C$_3$H$_2$}
\label{sec.cC3H2}
Two spectral features of the carbon chain c-C$_3$H$_2$ are detected. The one shown in Fig~\ref{fig.Integrated}, which is a combination of the $6_{1,6}-5_{0,5}$ and $6_{0,6}-5_{1,5}$ transitions, 
and a second feature centered at 217.940~GHz corresponding to the $5_{1,4}-4_{2,3}$ transition. However this feature is $\sim2$ times fainter and is mostly undetected in our maps.
This feature appears at the lowest velocities in the channel maps in Fig.~\ref{fig.channelcC3H2} (online).
Even at the positions where both features are detected, it is not possible to make a reliable estimate of the excitation temperature due to the limited dynamic range in the energies
involved in these transitions.

The emission of c-C$_3$H$_2$ is faint in the CND.
It is barely seen in both the integrated map (Fig.~\ref{fig.Integrated}) but integrated channel maps (Fig.~\ref{fig.channelcC3H2} online) show c-C$_3$H$_2$ to follow HC$_3$N.
We note, however, that we find some selected positions where c-C$_3$H$_2$ is detected but we find no trace of HC$_3$N (see Table~\ref{tab.GaussFits} online).

It has been claimed that c-C$_3$H$_2$ is mainly formed in gas phase by dissociative recombination of c-C$_3$H$_3^+$ \citep{Gerin2011} and is much less sensitive to photodissociation
than HC$_3$N.
This molecule has been observed to be very abundant in both diffuse clouds \citep{Lucas2000} and in photodissociation regions \citep[PDRs,][]{Teyssier2004}.
Moreover, in their study of different PDRs, \citet{Teyssier2004} found the emission c-C$_3$H$_2$ and other small hydrocarbons up to the edge of the PDR.
On the chemical basis we would then expect the emission of c-C$_3$H$_2$ to be more prominent all over the CND region.
In the Horsehead nebula, a column density ratio of c-C$_3$H$_2$/HC$_3$N$\sim25\pm12$ is found in the position where both species are detected.
Across the CND, we find integrated line ratios of c-C$_3$H$_2$/HC$_3$N$=0.4-2.4$.
Due to the large upper level energy difference between the transitions observed for  c-C$_3$H$_2$ and HC$_3$N, the column density ratio strongly depends on the excitation temperature.
We will assume a range of likely temperatures of $20-50$\,K, corresponding to dust temperature components fitted by \citet{Etxaluze2011}.
We find column density ratios of $0.8-5\times10^{-2}$ and $0.3-1.6$, for a temperature of 20 and 50~K, respectively.
Thus, independent of the temperature assumption, abundance ratios in the CND differ significantly from those of PDR such as the Horsehead nebula.
However, it is unclear whether this is due to an underabundance of c-C$_3$H$_2$ or due to the enhanced emission of HC$_3$N in the CND region.
Observations of the light carbon chain CCH, intimately related to c-C$_3$H$_2$ \citep{Teyssier2004,Gerin2011} might be a better candidate to study the distribution of the carbon chains
in the vicinity of Sgr~A$^*$.

\subsection{HC$_3$N: The Warm Dense Molecular Component}
\label{sec.HC3Ndensest}
Similar to SiO $5-4$, the HC$_3$N $25-24$ transition is likely tracing the densest molecular component, with a critical density of $\sim2\times10^6\rm cm^{-3}$ \citep{Wernli2007}.
Indeed, the emission of these two species appears to be coexistent and reaching its maximum emission towards the southern region of the CND (in both the southwest lobe and
the southern extension) and the northeast arm. Both species show the largest enhancement with respect to CN in the northeast arm.
We observed CN/HC$_3$N ratios of $\sim5-6$ in the southern structures, down to ratios of $\sim 2$ in the northeast arm.

However, the formation and destruction of HC$_3$N differs from that of SiO.
HC$_3$N is efficiently formed in the gas phase by neutral-neutral reactions involving CN and C$_2$H$_2$, which can be abundant in an UV irradiated environment \citep[][and references therein]{Meier2005}.
Additionally, HC$_3$N is the species with the highest photodissociation rate (Fig.~\ref{fig.photorate}) and is efficiently destroyed with reactions with C$^+$.
Therefore, in the heavily irradiated CND region, only within moderately shielded clouds \citep[$A_v>1-2$][]{Woodall2007}, the formation reaction of HC$_3$N will overcome the photodestruction rate.

Thus, the high$-J$ transition of HC$_3$N with an upper level energy of$\sim150$\,K is tracing the warmest dense molecular clumps, shielded from the photodissociation radiation from Sgr~A$^*$.
In the southwest lobe and extension the HC$_3$N is observed in the outer edge of the CN emitting region, where the UV radiation is attenuated.
HC$_3$N shows its brightest emission in the southern part of the northeast arm, where it shows the lowest ratio SiO/HC$_3$N$<1$.
However, though SiO also traces the dense gas material, the formation of HC$_3$N via CN explains that no HC$_3$N emission is observed in regions where CN is absent, as seen in the SiO maps.

\section{The spectra towards Sgr~A$^*$}
\begin{figure}[]
\centering
\hspace{-20pt}
\includegraphics[angle=-90,width=0.5\textwidth]{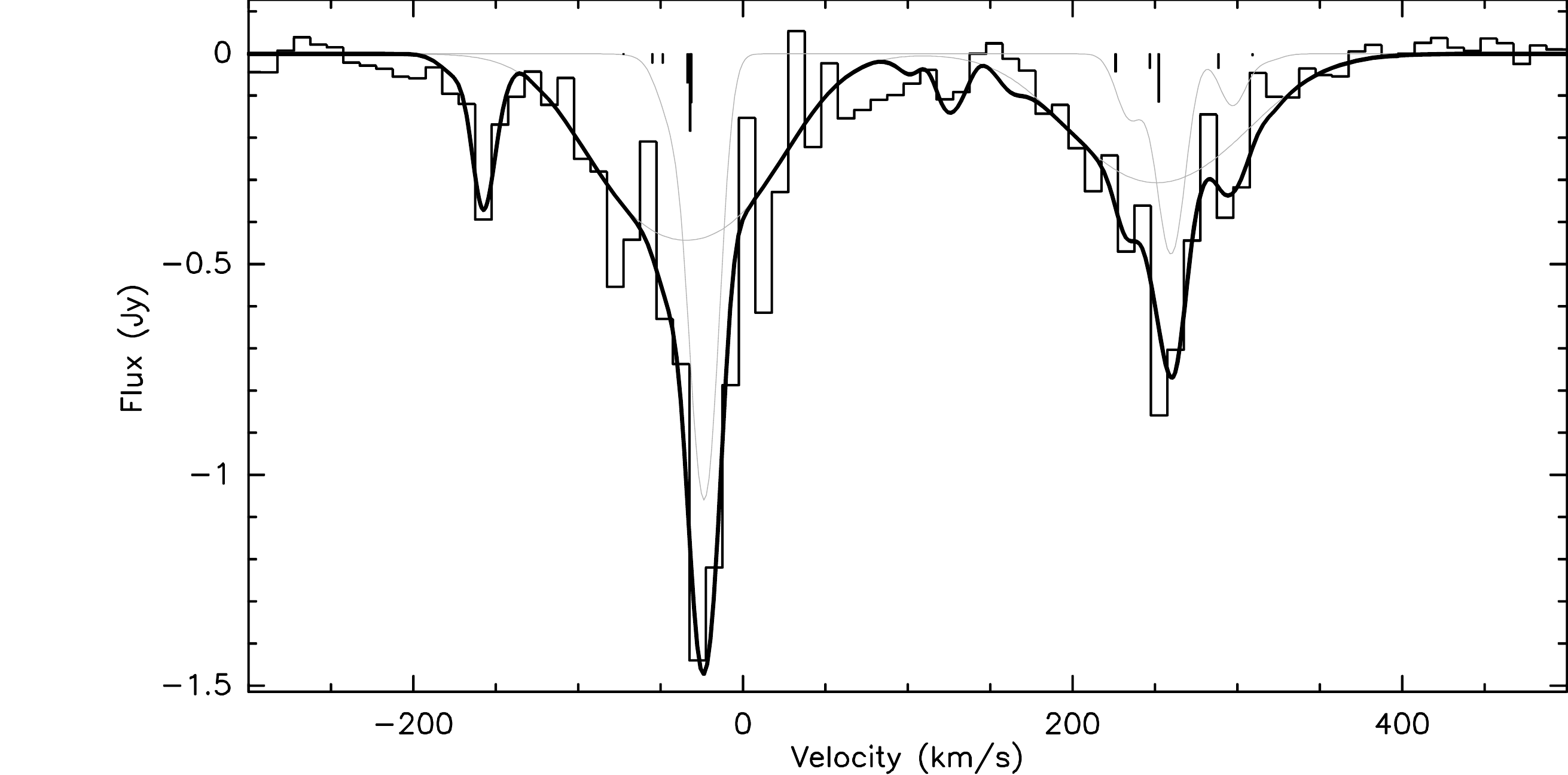}
\caption{Continuum subtracted CN $2-1$ spectra extracted from a $4''$ box around the Sgr~A$^*$ position. The thick black line represents the fit performed to the line profile
where three velocity components have been considered.
Two narrow ($\sim 20$ \kms) absorption features are fitted at $-157$ and $-23$~\kms\, and a \mbox{$\sim 100$~\kms\,} wide feature at $-32$~\kms.
The thin grey line shows the individual contribution of the narrow $-23$ ~\kms~ component and the wide component at $-32$~\kms.
Vertical lines show the position and relative intensities of each individual CN hyperfine structure transition.
The intensities and positions of these lines correspond to the fit results to the wider component. 
\label{fig.specSgrA}}
\end{figure}

The low-$J$ transitions of HCN show absorption spectral features towards the position of Sgr~A$^*$ \citep{Guesten1987,Wright2001,Christopher2005}.
In Fig.~\ref{fig.specSgrA} we show the CN absorption spectra towards the Sgr~A$^*$ position.

We have fitted a 3 component hyperfine group profile to the observed spectrum.
In Fig.~\ref{fig.specSgrA} we indicate the relative position and intensities of the CN hyperfine components as well as the fit results.
The two narrow features at $-157$~\kms\, and $-23$~\kms, with widths of 16 and 20~\kms, respectively,
had been previously identified as absorptions due to galactic spiral arms along the line of
sight at $-135$, $-50$ and $-30$~\kms\, \citep[see][]{Wright2001}.
Our $-23$~\kms~ feature is likely a composite of the extended local gas at 0~\kms~ \citep{Guesten1987} and the $-50$ and $-30$~\kms~ galactic arms.
The opacities derived from the fit for the main hyperfine transition are $\sim0.1$ and 0.9 for the $-157$~\kms\, and $-23$~\kms~ components, respectively.
Both narrow features are relatively optically thin.

On top of these narrow features we fit a $94$~\kms\, wide component centered at $-32$~\kms. Different from the narrow features, the opacity of
the main hyperfine transition is $\tau_{\rm CN}\sim3$.
This is likely the material closer to Sgr~A$^*$, observed in absorption in the line of sight to the non-thermal central source emission due to the
low-$J$ of the transition. Down to zero level the absorption extends from $\sim-150$~\kms\, to $\sim100$~\kms, wider than the absorption
observed from HCN or HCO$^+$ \citep{Christopher2005}.
Moreover, the residual absorption in the velocity range around -200 and 100~\kms\, might suggest an even wider ($>100$\kms) absorption component.

The absorption spectrum towards the central compact sources is particularly interesting to study the chemistry of the molecular gas in its line of sight.
However,regarding the other molecular species presented in this paper, none is detected either in absorption or emission towards the Sgr~A$^*$ position.
This absence reflects abundance deficiencies of these species towards the galactic arms and also in the inner molecular gas around Sgr~A$^*$.
The intense UV radiation from the central cluster is likely responsible for the photodissociation of most molecular species in the very
central region.

\section{CND clump kinematics}
\label{sec.kinematics}

\begin{figure*}[]
\centering
\includegraphics[width=0.9\textwidth]{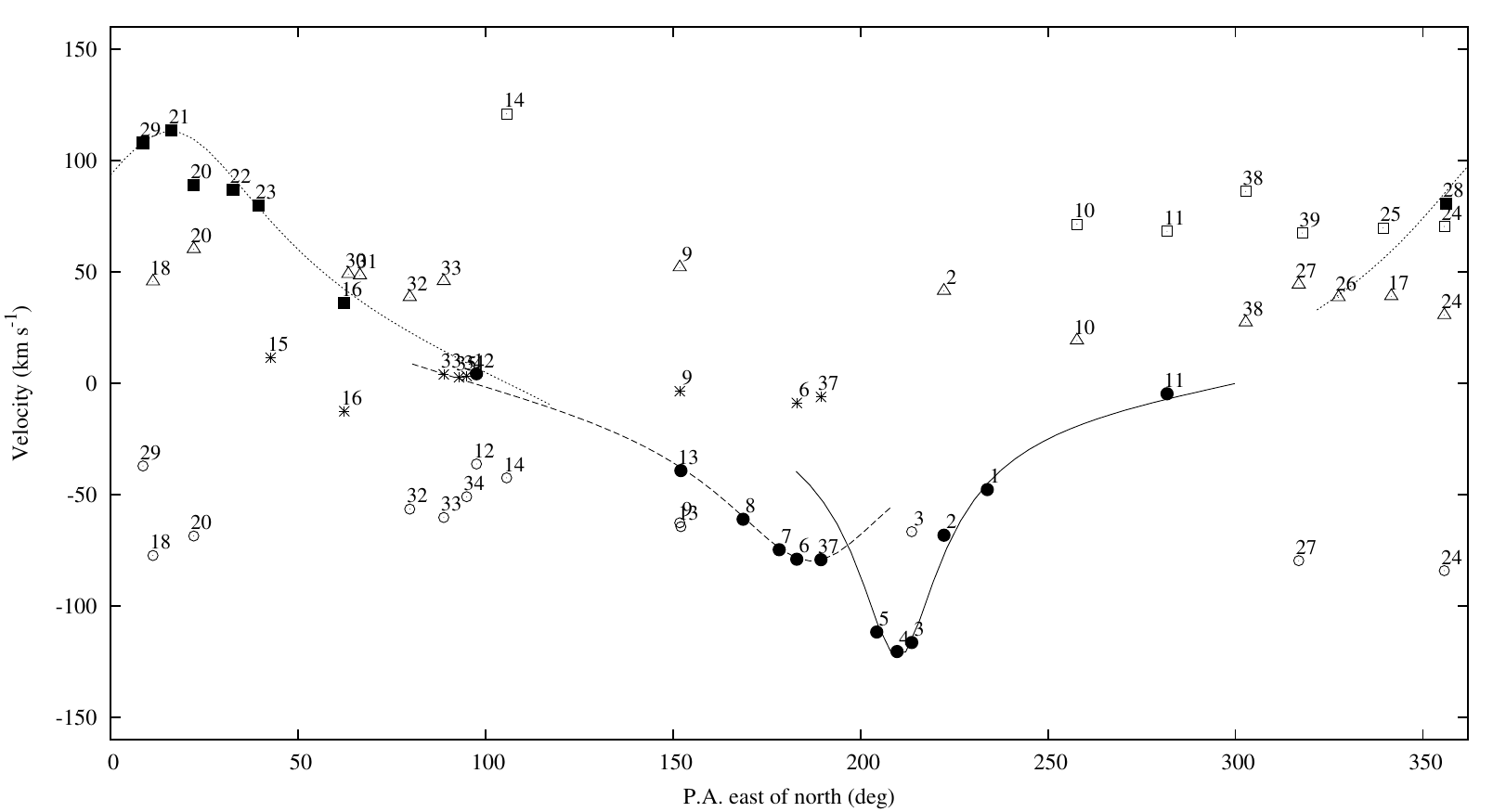}
\caption{Velocities of each of the CN molecular components fitted at each of the core positions labeled in Fig.~\ref{fig.Integrated} as a function of
the position angle on the sky east of north as measured from Sgr~A$^*$.
Each marker is labeled with the corresponding position where the velocity was measured.
The different symbols represent different groups of cores/components: southern CND (filled circles), northern CND (filled squares), 70 \kms sources (open squares), 50 \kms sources (open triangles),
0 \kms sources (stars), and negative velocities sources (open circles).
The overlaid lines represent rotating ring models fitted to the core velocities in the southwest lobe (solid line), southern extension (dashed line), and northeast lobe (dotted line).
See Sect.~\ref{sec.kinematics} for details on the sources and models represented.
\label{fig.VeloVsPA}}
\end{figure*}

In this section we aim to analyze the kinematics of all molecular emission peaks selected and labeled in Fig.~\ref{fig.Integrated} taking into account all the velocity components fitted to each
individual CN spectrum as compiled in Table~\ref{tab.CNvs13CN} (online).
We selected CN for this study as it is the only species that shows emission in all selected positions.
A total of 66 velocity components have been identified and fitted towards the 38 selected molecular clumps (where we do not include the central Sgr~A$^*$ position).

Fig.~\ref{fig.VeloVsPA} shows the measured velocities as a function of the position angle (P.A.) east of north as measured from the position of Sgr~A$^*$.
Similar plots have been presented by \citet{Guesten1987} and \citet{Jackson1993} where they plotted the velocity centroids along the CND.
However, in this work we take into account the individual velocity components at each selected position 
so that we can disentangle in velocity space the different molecular clumps in the line of sight.
At each position we find molecular velocity components both related to the different CND rotating structures and kinematically detached from it.
Only $29\%$ of all the velocity components are associated with rotating structures around Sgr~A$^*$ (see Sect.~\ref{sect.CNDrings}), while the remaining $\sim70\%$ belong to the overlapping
GMCs or likely foreground material along the line of sight.
However, these clumps in rotating structures represent $67\%$ in mass of molecular gas.
This estimate, based on the fraction of CN integrated intensity in the clumps, turns into a $70\%-75\%$ when corrected by the opacity of CN, mostly affecting the clumps associated with the 
rotating structures (Sect.~\ref{sect.1213C}).

\subsection{CND molecular components}
\label{sect.CNDrings}
The CND is formed by distinct rotating structures rather than a single structure \citep{Jackson1993,Wright2001}.
Our observations seem to prove the presence of three kinematically distinct structures in the CND.
We have assumed a simple rotating ring model, similar to that used by \citet{Guesten1987}, to describe the rotation around Sgr~A$^*$ of each of these structures.
Even though these structures are not complete rings, it is a reasonable approximation to describe the rotation in small position angle (P.A.) ranges.

The southern molecular structure of the CND (filled circles in Fig.~\ref{fig.VeloVsPA}) cannot be described by a single ring model. This fact could be observed in the data from \citet{Guesten1987}.
However, it is possible to fit two different ring models with inclinations of $\sim80^\circ$ (solid line) and $\sim68^\circ$ (dashed line) for the southwest lobe and the southern extension,
respectively. The peak rotation velocity ($v_{\rm rot}/sin(i)$) for these model were 125 and 86 \kms, respectively.
We observe that the southwest lobe model might extend northwards up to position 11, but would not fit position 10 (see Fig.~\ref{fig.Integrated}).
On the other hand, the southern extension model might extend eastwards to positions 13 and 12, and towards the west to position 37, which is located in the connecting point between the southern extension
and the southwest lobe structures.
These results suggest that, though close in projection, both southern structures do not belong to the same rotating structure, and indeed might be fed by different streamers (see Sect.~\ref{sec.SiOshocks}).

Similarly, we can fit a ring model to the northeast lobe components (filled squares) with an inclination of $\sim65^\circ$ and a peak velocity of 125 \kms.
This structure might start in positions 28 and 29, where H$_2$CO shows a peak of emission, and would continue southwards through positions 21 to 23.
The northeast extension (position 20, also in filled squares) is not sampled with enough points in our study, but it is clearly detached from the kinematics of the northeast lobe.

\subsection{Other molecular components}
\label{sect.CNDnonrings}
The rest of the molecular components sampled do not seem to be associated with any evident rotating structure but rather with larger scale structures placed on top of the CND.

We find a number of component at velocities of $\sim 50$~\kms\, (open triangles). This positions are located mostly towards the north-east of the maps and are therefore likely
related to gas condensations in the 50 \kms~ giant molecular cloud \citep[see SiO and CS single-dish maps by][]{Amo-Baladr'on2011}.
Mostly towards the north-western parts of the CND we find a molecular component at $\sim70$~\kms. As observed in the CS maps by \citet{Amo-Baladr'on2011} this could be an extension of
the 50~\kms\, giant molecular cloud.
At velocities of $\sim0$~\kms\, we find a number of components (stars in Fig.~\ref{fig.VeloVsPA}). These positions are mostly located towards the south-east of our maps and could be associated with the
20~\kms\, GMC and the molecular bridge connecting to the 50~\kms\, GMC.
Finally we observed molecular gas at negative velocities (open circles). Mostly located towards the north-east of the CND, they are likely related to different molecular cloud complexes along the
line of sight. However, no evident connection is found with the CS maps by \citet{Amo-Baladr'on2011}.

Though the general trends described here for CN apply to all species, some significant differences are found in the velocity component fitted in the other detected species (see Table~\ref{tab.GaussFits} online).
In particular, the emitting region east of the southern extension (Sect.~\ref{sec.H2COshell}), where no CN is detected, shows emission peaks at the velocities of $\sim-9$ and 11~\kms~ (position 36 in Table~\ref{tab.GaussFits} online)
consistent with the molecular gas being associated with both foreground clouds and the 20~\kms~GMC.

\section{CND structure and interaction with the mini-spiral}
\begin{figure}[]
\centering
\includegraphics[angle=-90,width=0.46\textwidth]{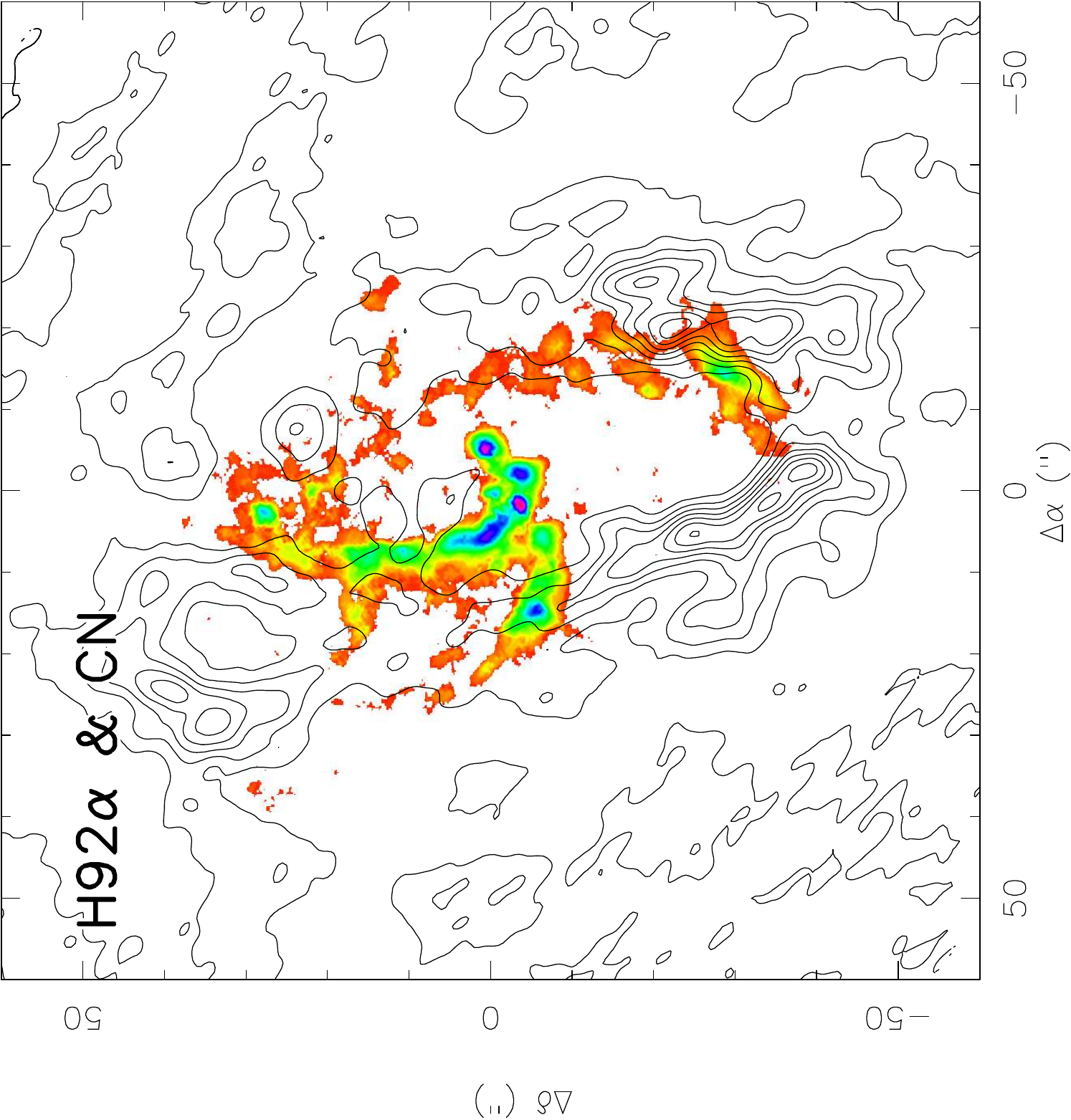}
\includegraphics[angle=-90,width=0.46\textwidth]{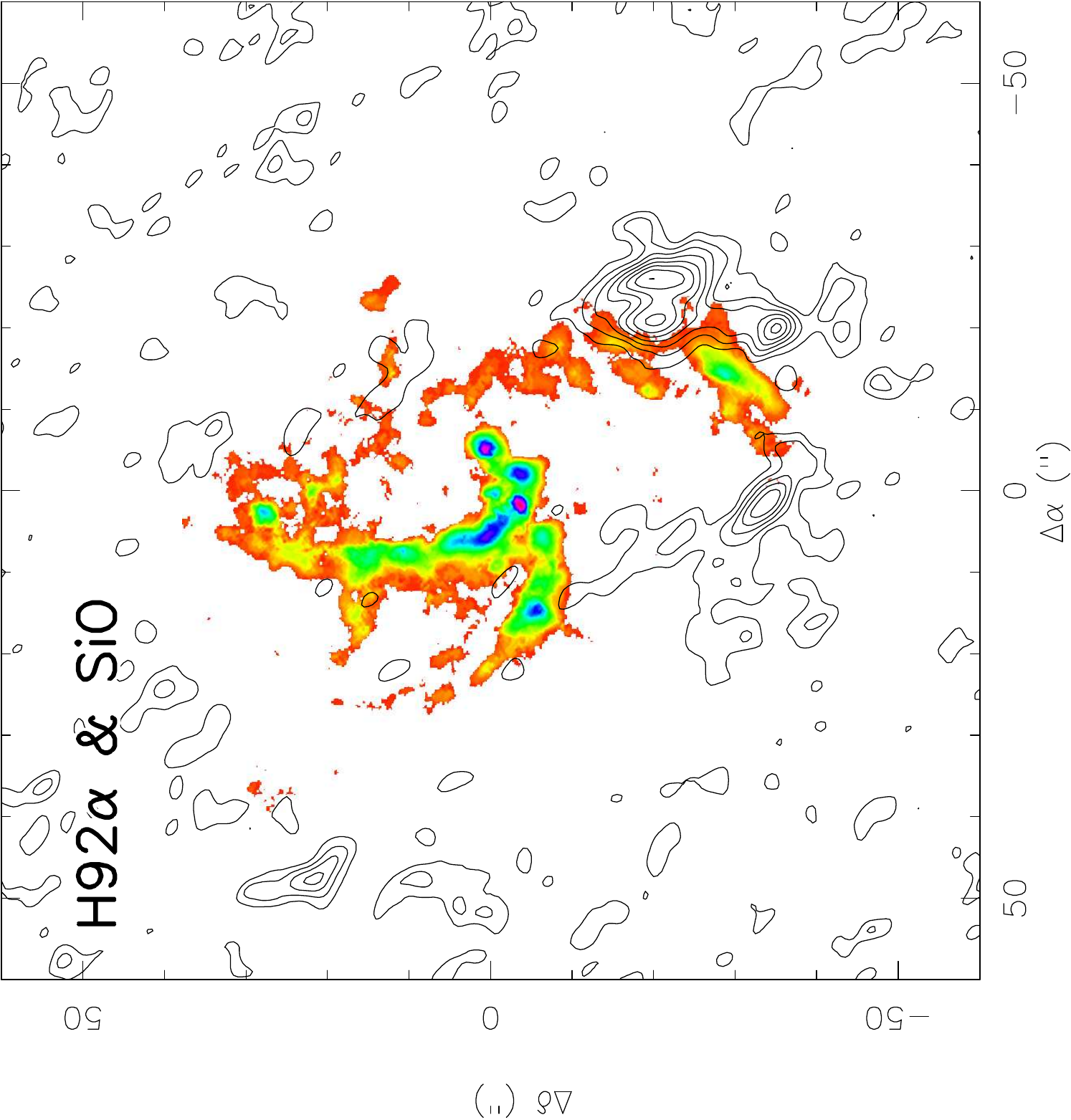}
\caption{
({\it UPPER}) Comparison between the H92$\alpha$ integrated emission \citep[color;][]{Roberts1993,Zhao2009} and the CN integrated emission (contours).
CN contour levels are the same as in Fig.~\ref{fig.HCNvsCN}.
({\it LOWER}) Comparison between the H92$\alpha$ integrated emission \citep[color;][]{Roberts1993,Zhao2009} and the SiO integrated emission (contours).
SiO contour levels are the same as in Fig.~\ref{fig.Integrated}.
\label{fig.H92avsCNandSiO}}
\end{figure}

The association of the gas in the CND with the ionized mini-spiral structure has been long discussed
\citep[][and references therein]{Christopher2005}.
The inner edge of the CND in the southwest lobe appears to be tightly outlined by the western ionized arc.
This correlation has led to the idea of the western arc being the ionized boundary of the CND where
the strong pervasive UV field from the central cluster is responsible for the heating of both the CND \citep{Mart'in2008,Amo-Baladr'on2011} and the mini-spiral
structure \citep{Latvakoski1999}.
Such an association was kinematically confirmed by \citet{Christopher2005}.
In Fig.~\ref{fig.H92avsCNandSiO} we compared our CN observations with the ionized gas traced by the H92$\alpha$ emission from \citet{Roberts1993}.
We observe that not only the emission is tracing the inner edge of the southwest lobe, but the H92$\alpha$ appears
to trace the inner side facing Sgr~A$^*$ of the CN emitting region.
Similarly,
the emission peaks of SiO in the southwest lobe are observed in the emission gaps of H92$\alpha$ (lower panel in Fig.~\ref{fig.H92avsCNandSiO}), 
as if the ionized gas would be tracing the inner boundaries of the SiO emitting gas.

\citet{Christopher2005} noticed a possible connection of the northern arm of the mini-spiral and the 
northeastern extension of the CND.
Our CN observations, in agreement with the HCN $4-3$ maps by \citet{Montero-Castano2009}, show a more prominent emission
in the northeast extension connecting to a faint but continuous eastern CND molecular emission.
Again, the H92$\alpha$ northern arm appears to be tracing the inner edge of the northeast extension of the CND which might
show a similar association as that claimed for the western arc.

We also observe an anti correlation between the ionized gas and the molecular material. 
The brightest northern arc outlines the inner edge of the faint molecular gas northeast extension,
while the fainter western arc traces the edge of the bright molecular southwest lobe.
This could be related to
the age of each of the CND structures and the history of molecular fueling from the outer molecular clouds into the CND.
In this scenario, the northeast extension would have been mostly ionized in the northern arm and little molecular gas would be left, while the southwest lobe, mostly molecular, would
start getting ionized as it approaches the central region.

\begin{figure}[]
\centering
\includegraphics[angle=-90,width=0.46\textwidth]{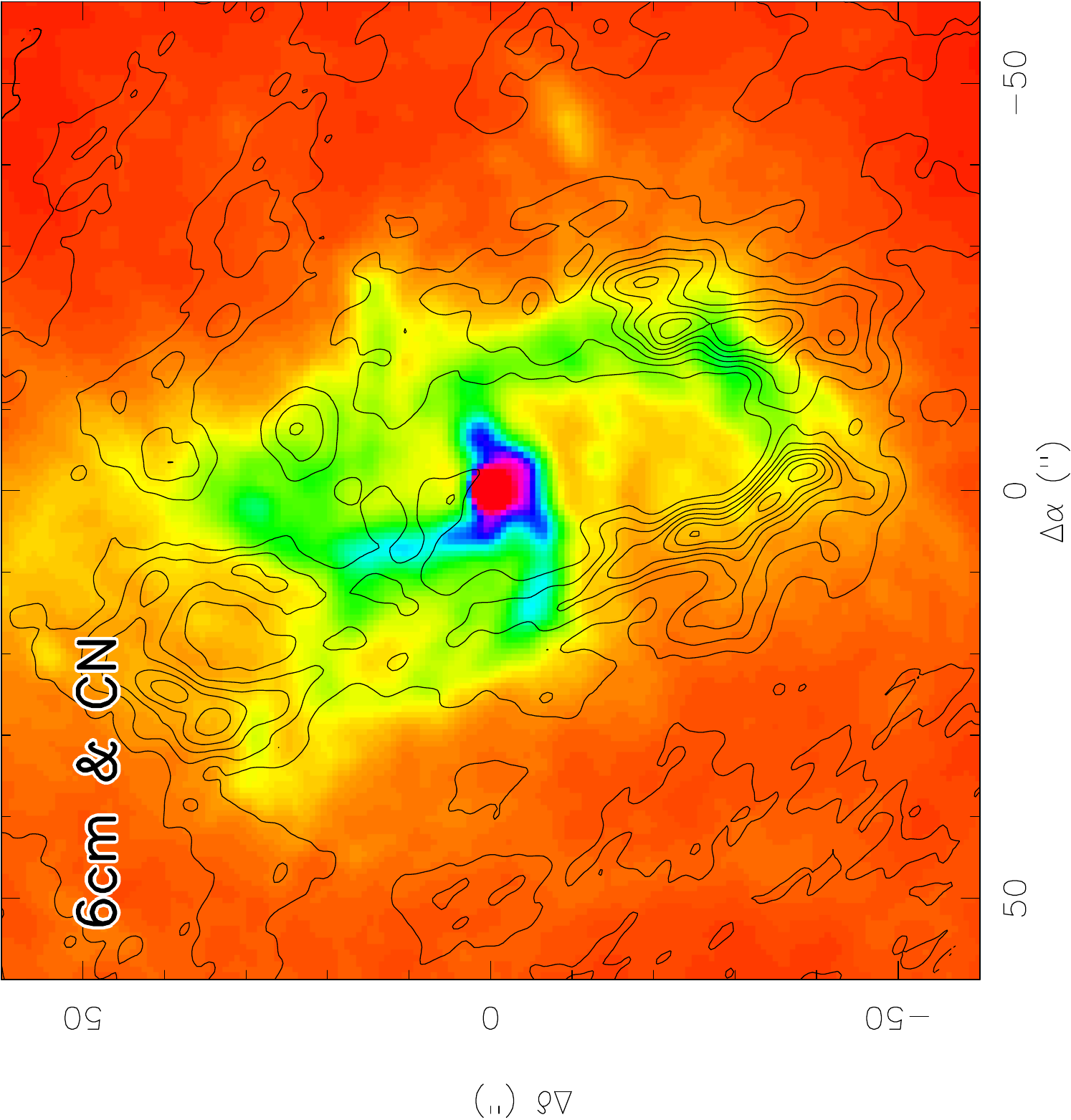}
\caption{Comparison between the 6~cm continuum emission \citep[color;][]{Yusef-Zadeh1987} and the CN integrated emission (contours).
CN contour levels are the same as in Fig.~\ref{fig.HCNvsCN}.
Similar to the comparison with H92$\alpha$ in Fig.~\ref{fig.H92avsCNandSiO},
the association of the northern arm and the western arc with the northern molecular extension and the southwest lobe can be appreciated.
The 6~cm shows the clear extension of the eastern arm northwards where it meets the bright molecular northeast lobe.
\label{fig.6cmvsCN}}
\end{figure}

\begin{figure*}[]
\centering
\includegraphics[width=0.85\textwidth,angle=-90]{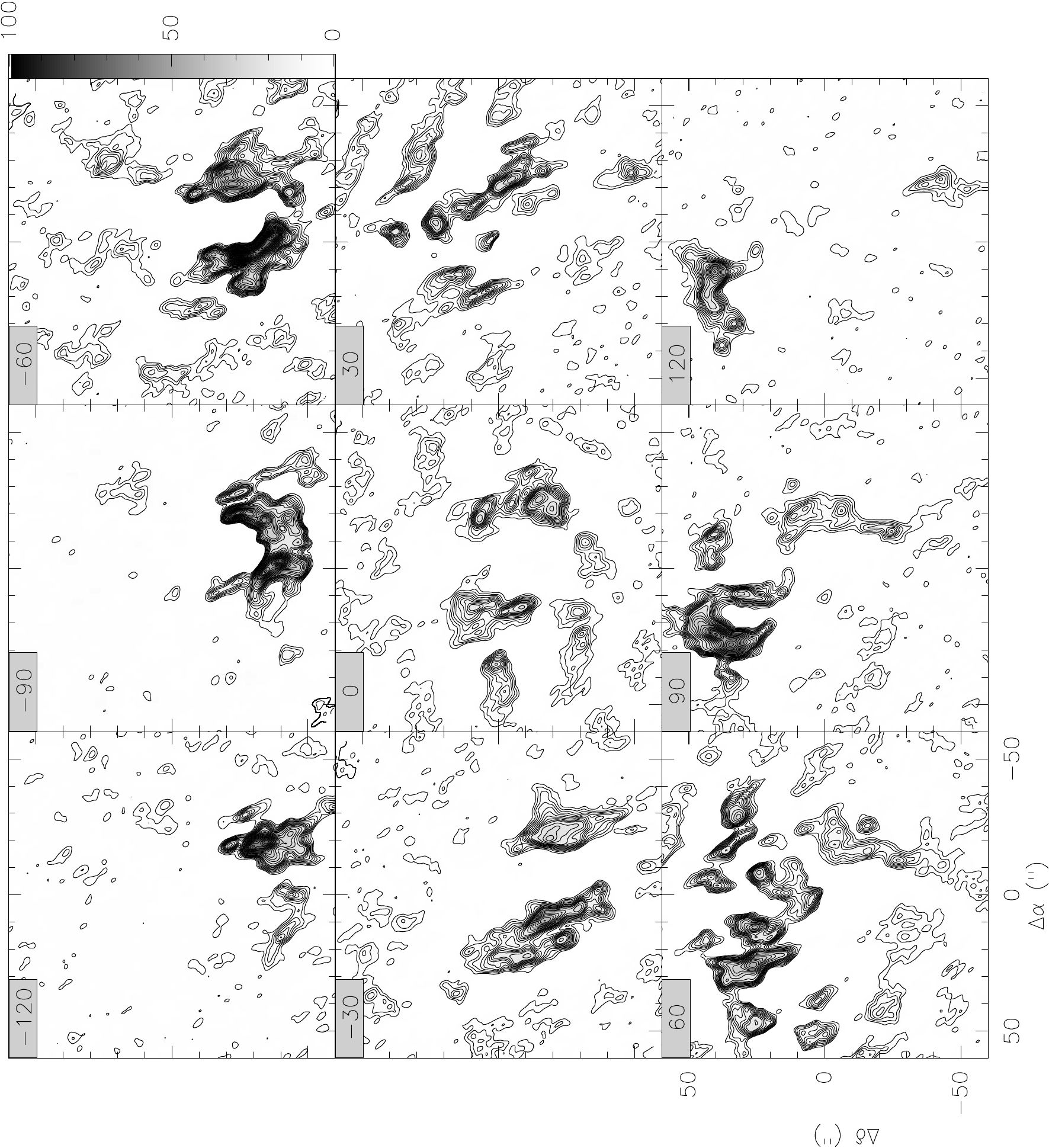}
\caption{Integrated intensity maps of CN $2-1$ in velocity bins of 30\kms.  Central velocity of each velocity range is displayed in the upper left corner of each panel. First 10 contours are in steps of 3$\sigma$,
with $\sigma=0.55\,\rm Jy\,beam^{-1}$\,\kms, and subsequently in steps of $6\sigma$.
\label{fig.channelCN}}
\end{figure*}

\begin{figure*}[]
\centering
\includegraphics[width=0.85\textwidth,angle=-90]{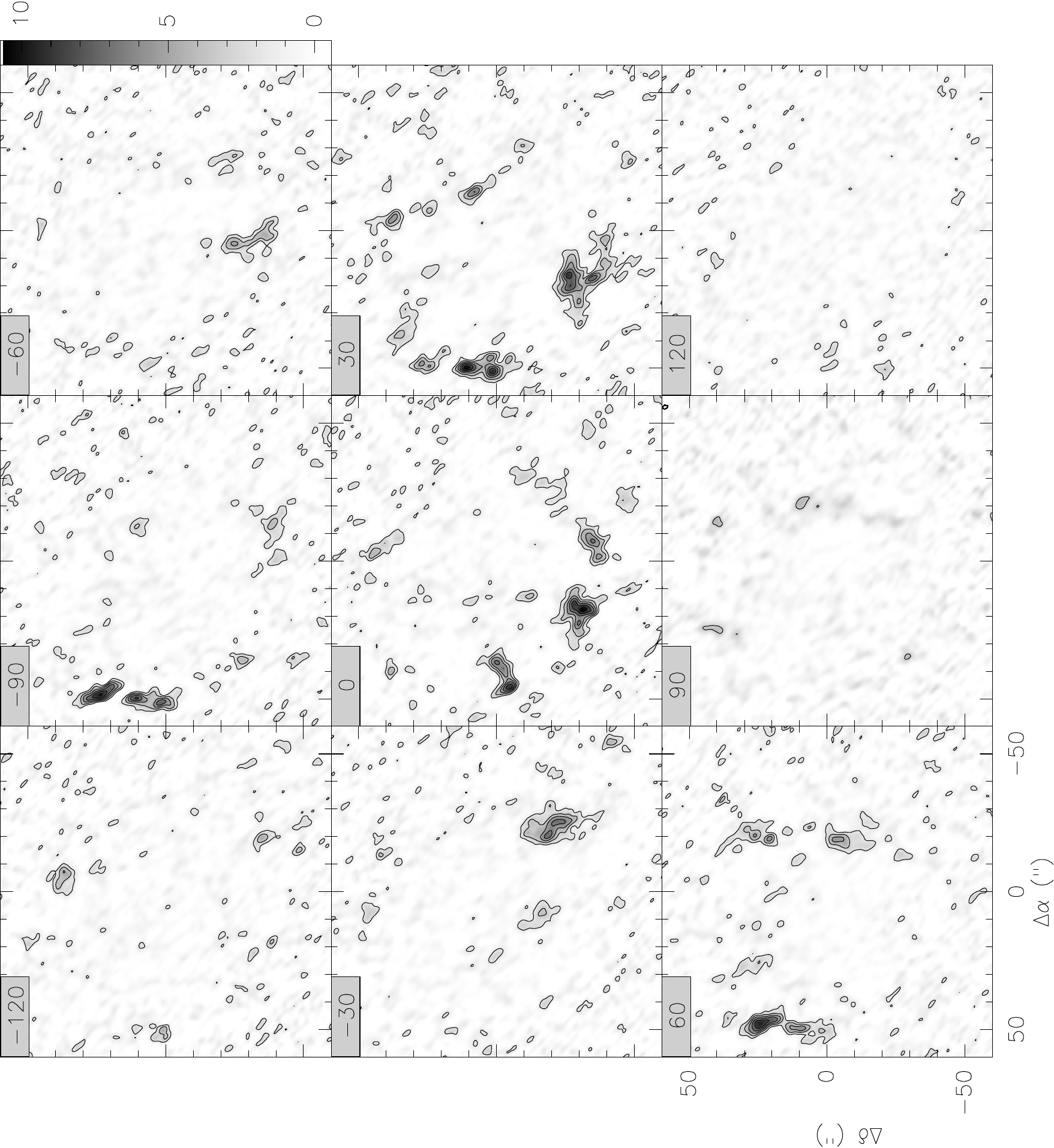}
\caption{Integrated intensity maps of H$_2$CO $3_{0,3}-2_{0,2}$ in velocity bins of 30\kms. Labels and contour steps as in Fig.~\ref{fig.channelCN}.
Note that the observed emission toward the north and east in the $-120$ and $-90$~\kms~ maps correspond to the high velocity emission of HC$_3$N $24-23$ (\@ 218.324 GHz)
detected at the edge of the observed band (Sect.~\ref{sect.results}).
\label{fig.channelH2CO}}
\end{figure*}

\begin{figure*}[]
\centering
\includegraphics[width=0.85\textwidth,angle=-90]{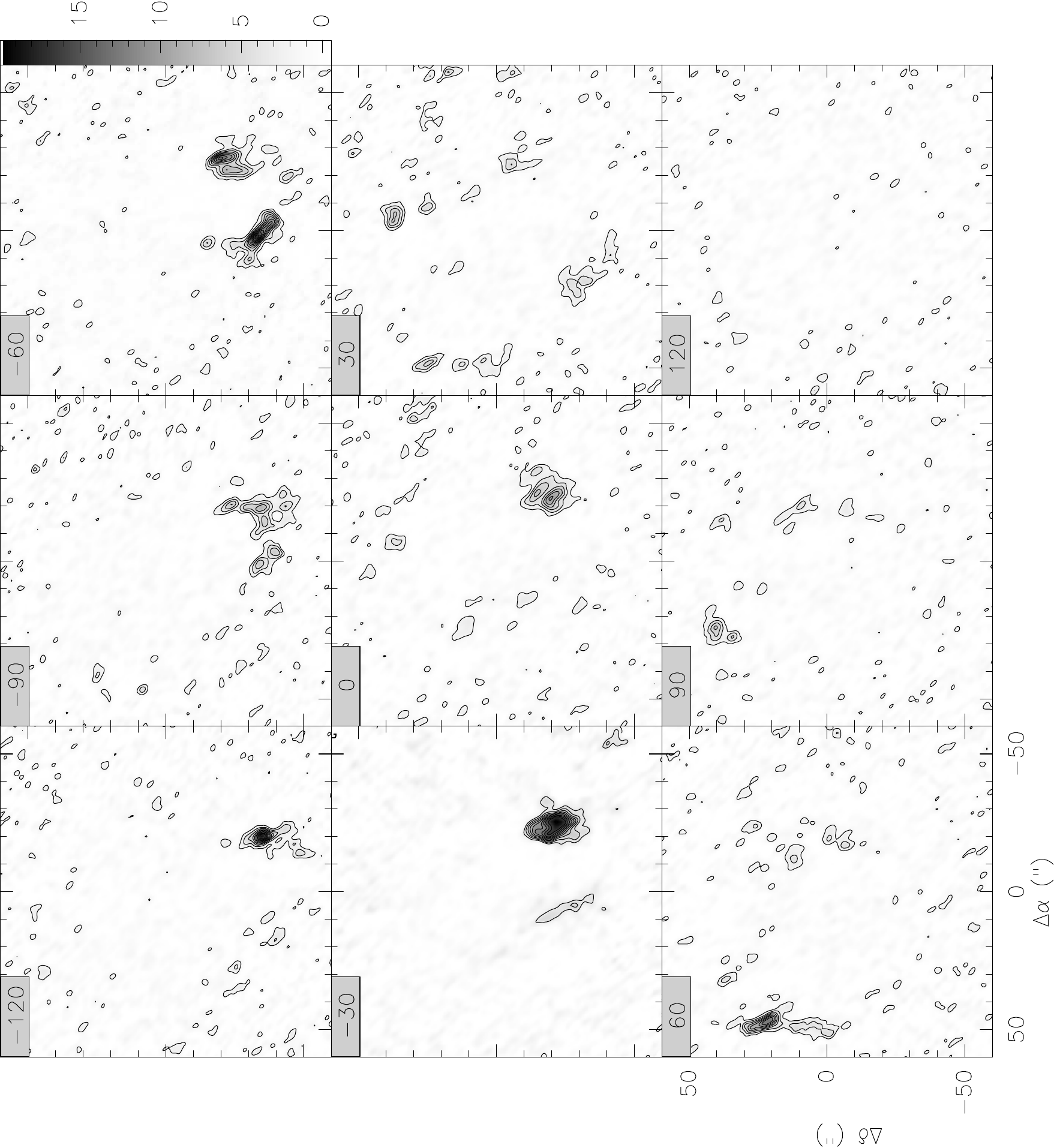}
\caption{Integrated intensity maps of SiO $5-4$ in velocity bins of 30\kms. Labels and contour steps as in Fig.~\ref{fig.channelCN}.
\label{fig.channelSiO}}
\end{figure*}

\begin{figure*}[]
\centering
\includegraphics[width=0.85\textwidth,angle=-90]{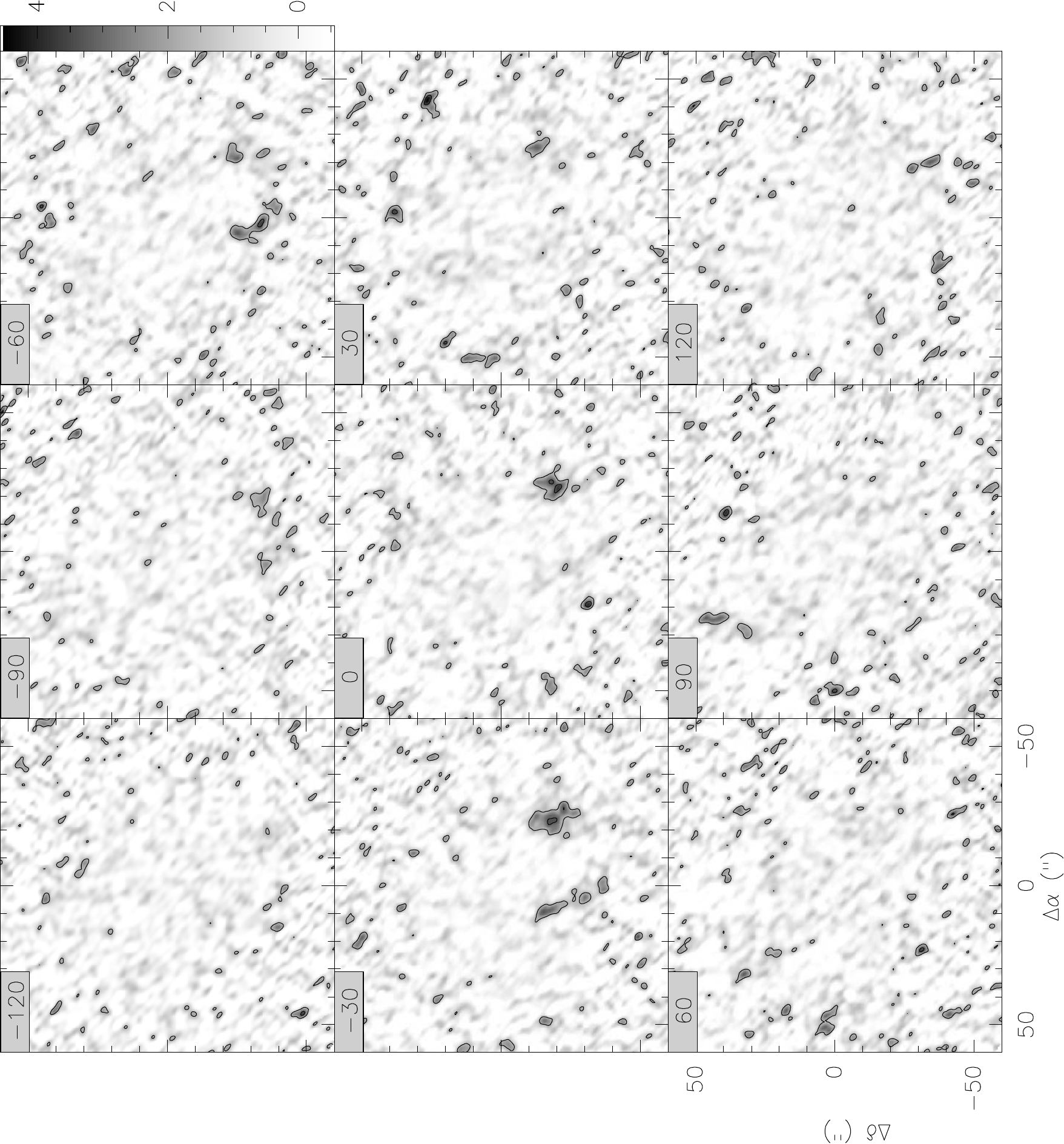}
\caption{Integrated intensity maps of $^{13}$CN $2-1$ in velocity bins of 30\kms. Labels and contour steps as in Fig.~\ref{fig.channelCN}.
\label{fig.channel13CN}}
\end{figure*}

\begin{figure*}[]
\centering
\includegraphics[width=0.85\textwidth,angle=-90]{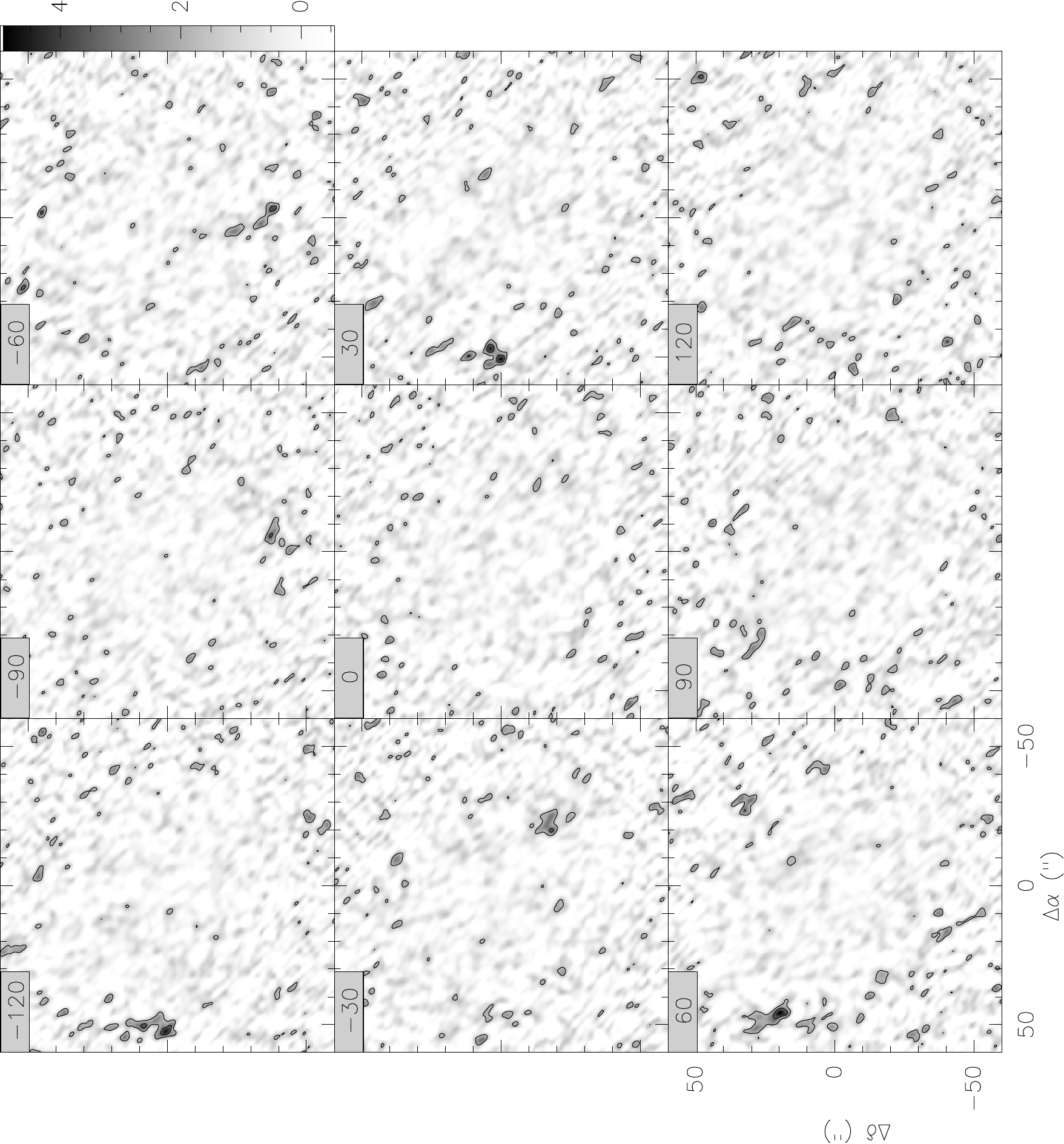}
\caption{Integrated intensity maps of $\rm c-C_3H_2$ $6_{1,6}-5_{0,5}$ in velocity bins of 30\kms. Labels and contour steps as in Fig.~\ref{fig.channelCN}.
Note that the observed emission toward the east in the $-120$~\kms map correspond to the high velocity emission of the $5_{1,4}-4_{2,3}$ transition of $\rm c-C_3H_2$
(\@ 217.940 GHz, see Sect.~\ref{sect.results}).
\label{fig.channelcC3H2}}
\end{figure*}

\begin{figure*}[]
\centering
\includegraphics[width=0.85\textwidth,angle=-90]{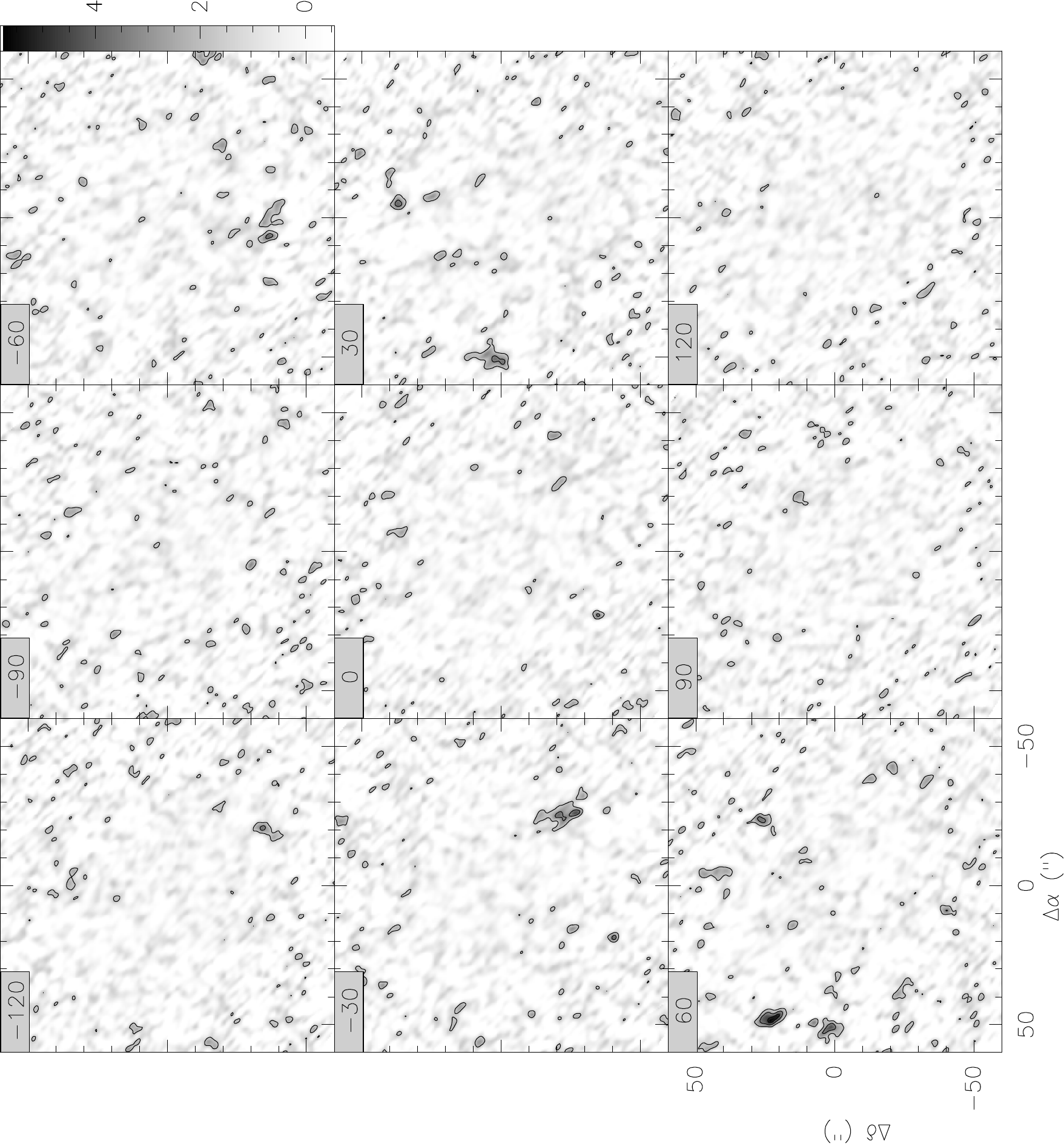}
\caption{Integrated intensity maps HC$_3$N $25-24$. Labels and contour steps as in Fig.~\ref{fig.channelCN}.
\label{fig.channelHC3N}}
\end{figure*}

\begin{figure*}[]
\centering
\includegraphics[width=0.85\textwidth,angle=-90]{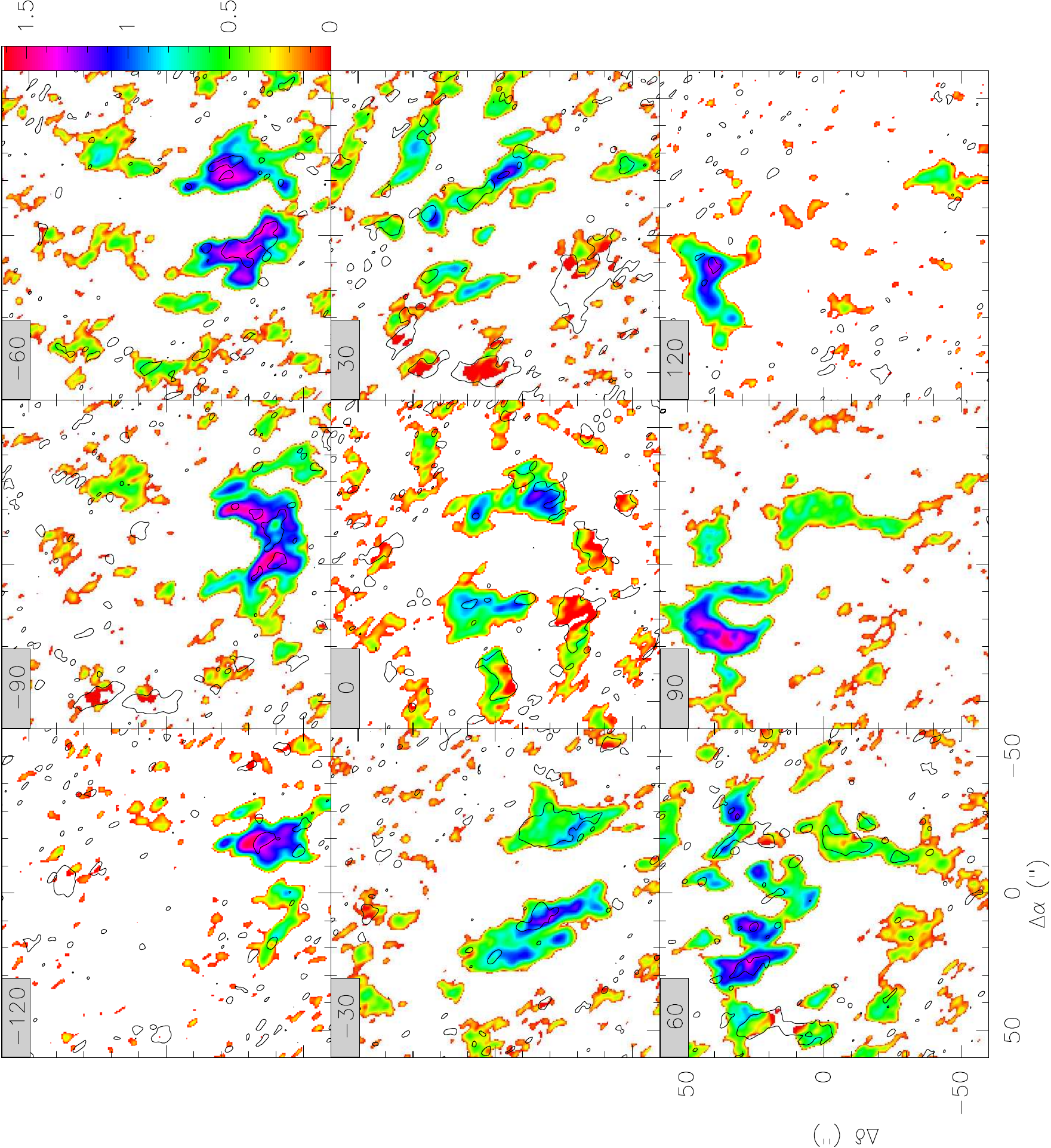}
\caption{Integrated CN/H$_2$CO line ratio in 30~\kms~velocity bin maps. Color scale is logarithmic. Ratio has been computed in all pixels where CN emission is detected above $3\sigma$.
The contour indicate the regions where H$_2$CO has been detected ($>3\sigma$). Therefore, values outside these contours represent lower limits to the line ratio
where a constant $3\sigma$ value has been assumed as the H$_2$CO upper detection limit.
\label{fig.channelCNvsH2CO}}
\end{figure*}

\begin{figure*}[]
\centering
\includegraphics[width=0.85\textwidth,angle=-90]{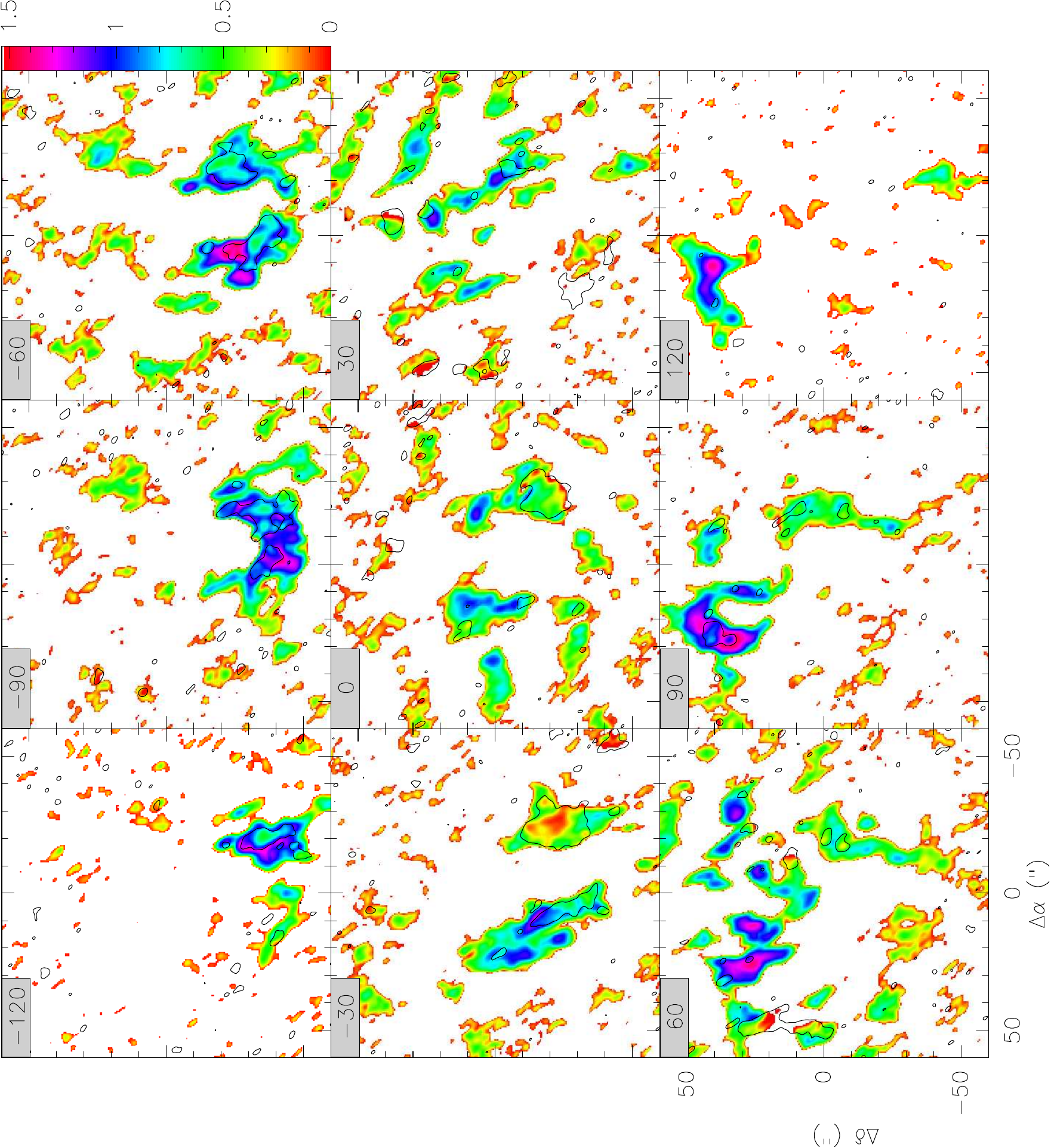}
\caption{Integrated CN/SiO line ratio in 30~\kms~velocity bin maps. Ratio has been calculated as in Fig.~\ref{fig.channelCNvsH2CO} where the contour enclose the regions
where SiO is detected above a $>3\sigma$ level.
\label{fig.channelCNvsSiO}}
\end{figure*}

\begin{figure*}[]
\centering
\includegraphics[width=0.85\textwidth,angle=-90]{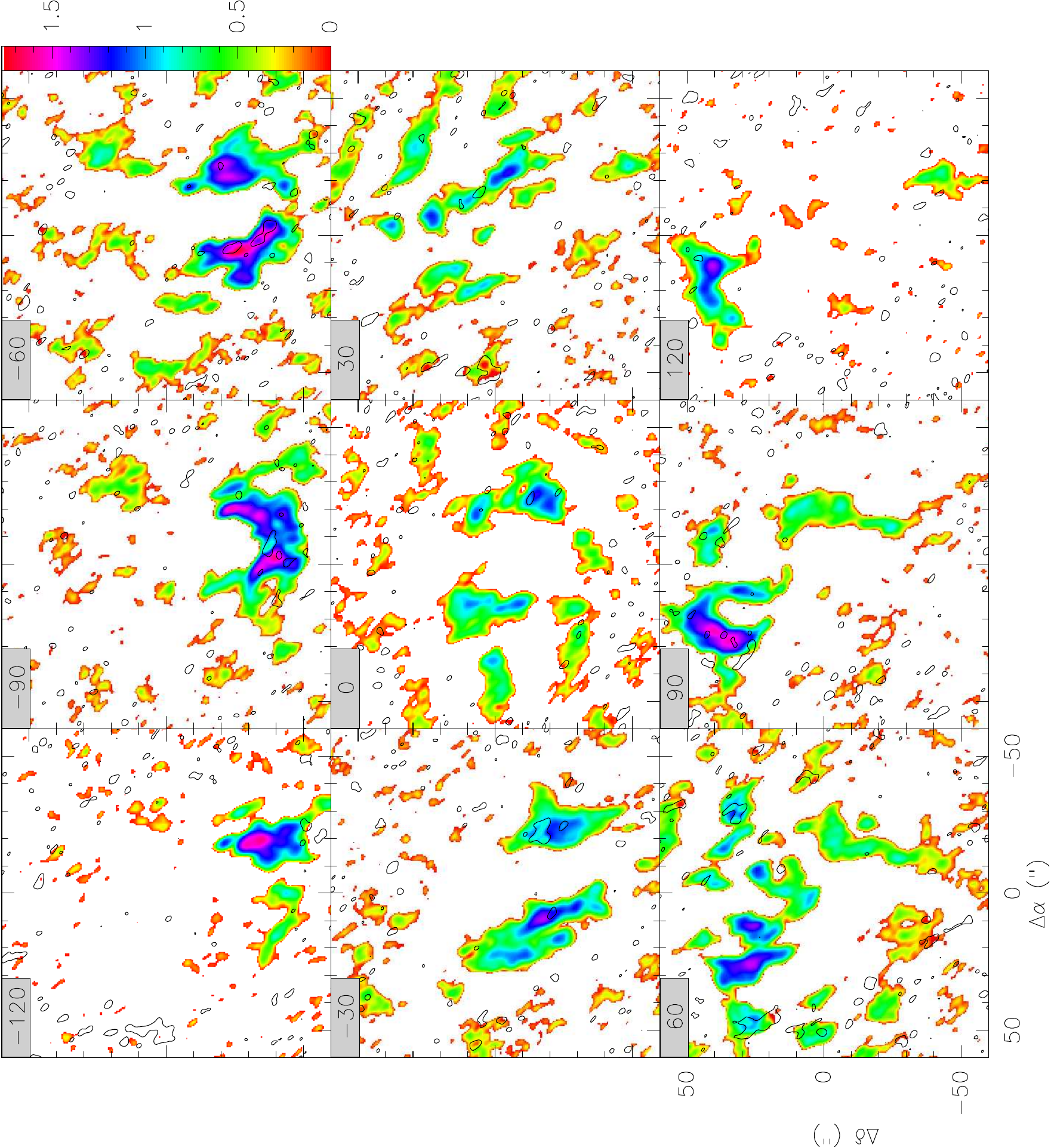}
\caption{Integrated CN/c-C$_3$H$_2$ line ratio in 30~\kms~velocity bin maps. Ratio has been calculated as in Fig.~\ref{fig.channelCNvsH2CO} where the contour enclose the regions
where c-C$_3$H$_2$ is detected above a $>3\sigma$ level.
\label{fig.channelCNvsC3H2}}
\end{figure*}

\begin{figure*}[]
\centering
\includegraphics[width=0.85\textwidth,angle=-90]{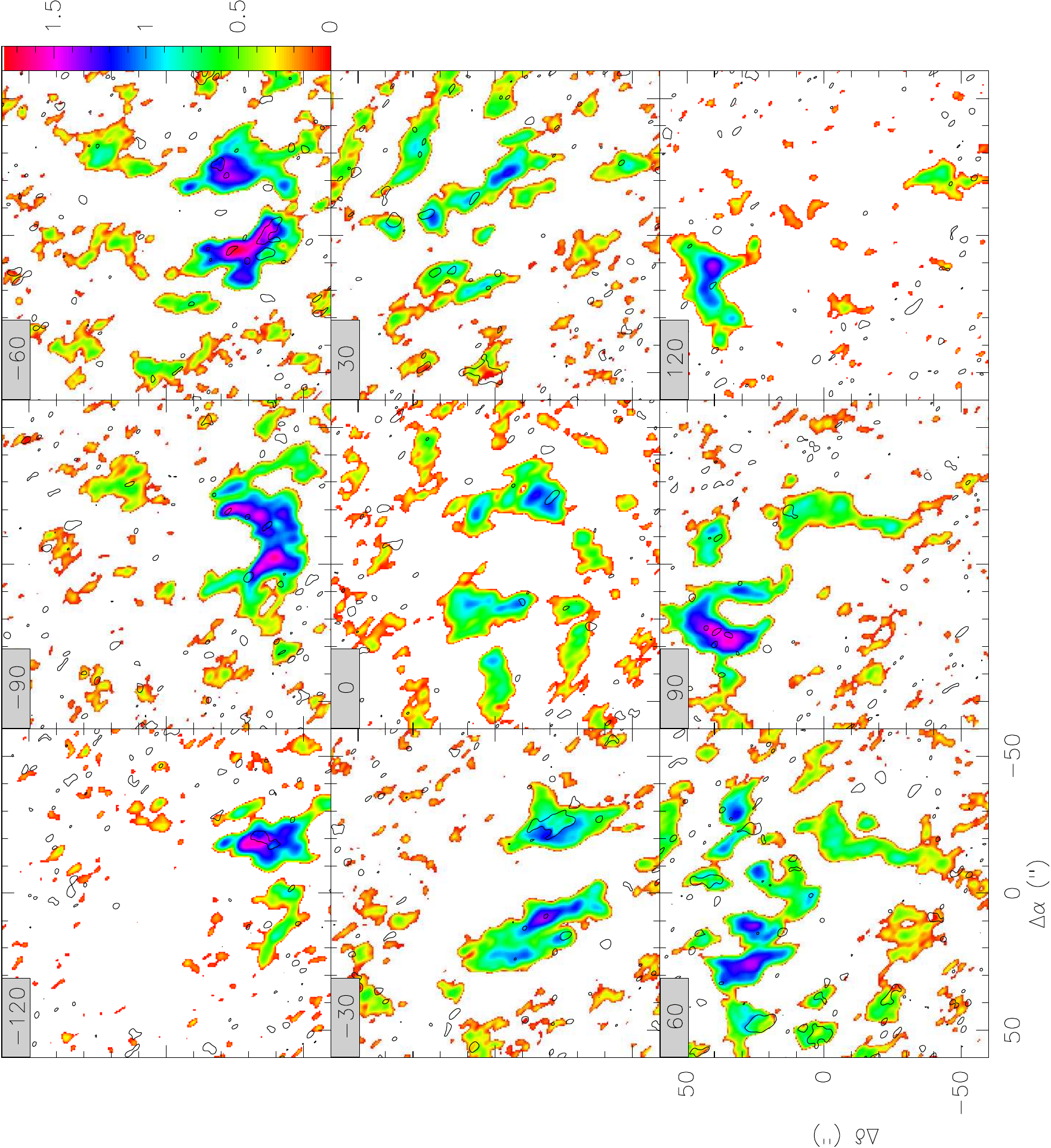}
\caption{Integrated CN/HC$_3$N line ratio in 30~\kms~velocity bin maps. Ratio has been calculated as in Fig.~\ref{fig.channelCNvsH2CO} where the contour enclose the regions
where HC$_3$N is detected above a $>3\sigma$ level.
\label{fig.channelCNvsHC3N}}
\end{figure*}

\begin{figure*}[]
\centering
\includegraphics[width=0.85\textwidth,angle=-90]{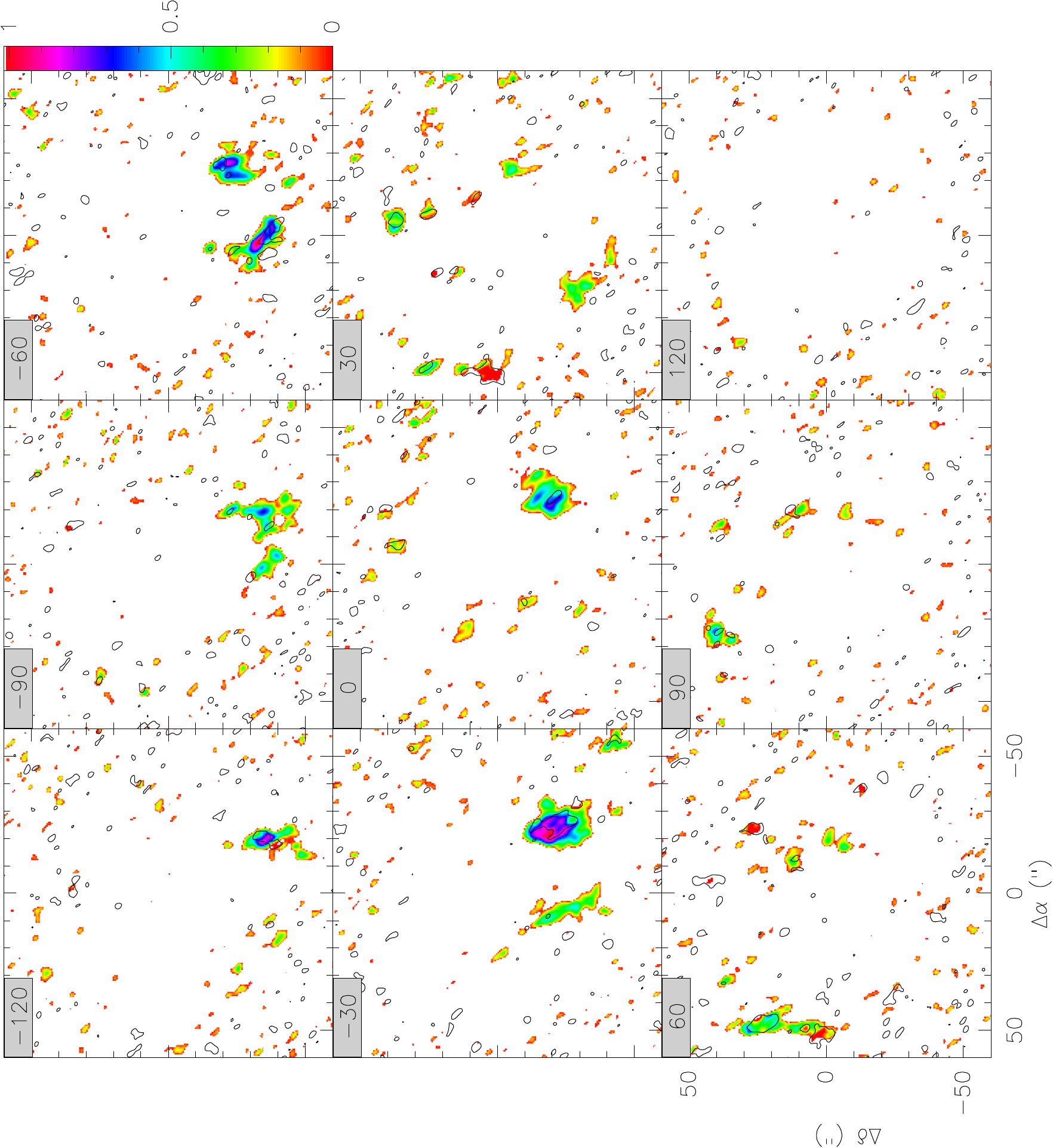}
\caption{Integrated SiO/HC$_3$N line ratio in 30~\kms~velocity bin maps. Ratio has been calculated in all pixels where SiO emission is detected above $3\sigma$.
Similar to Fig.~\ref{fig.channelCNvsH2CO}, the contour enclose the regions where HC$_3$N is detected above a $>3\sigma$ level.
\label{fig.channelSiOvsHC3N}}
\end{figure*}

The third mini-spiral arc, the eastern arm, has no obvious association with the molecular material.
The low resolution maps of HCN by \citet{Marshall1995} suggested this arc extends northwards likely towards the north lobe.
Indeed, at the position of $\sim(25'',30'')$, the eastern arm traced by H92$\alpha$ emission meets the northeast molecular lobe (Upper panel in Fig~\ref{fig.H92avsCNandSiO}).
Measured velocities from H92$\alpha$ at this position \citep[positions E1 and E2 in][]{Zhao2009}, match those in the northeast molecular lobe (positions 22 and 23 in Fig.~\ref{fig.VeloVsPA}).
Similarly, the eastern ionized arm extends northward as seen in the 6~cm radio maps \citep[Fig~\ref{fig.6cmvsCN},][]{Yusef-Zadeh1987}.
Though this arm appears to be crossing the eastern extension of the CND, there are no evidences of perturbation of the molecular material, which indicates that both features are not in the same plane.
Dust emission modeling \citep{Latvakoski1999} and H92$\alpha$ kinematics \citep{Zhao2009} showed the eastern arm must be significantly inclined with respect to the eastern CND.



\section{Summary}
In this paper we present high resolution maps of the CN, $^{13}$CN, H$_2$CO, SiO, c-C$_3$H$_2$ and HC$_3$N emission towards the central $\sim4$~pc around Sgr~A$^*$.
These observations provide a chemical picture of the molecular material in the surroundings of the central super-massive black hole and how the chemically resolved
structures relate to the different morphological and kinematical structures.
Each of the observed species appears to be tracing different molecular gas structures
which are intimately linked to the different heating processes affecting the molecular gas in this region.

Among them, CN is the only species showing emission in all previously described molecular morphological features in the CND, and it shows a similar
distribution to that from available HCN and CS emission maps \citep{Christopher2005,Montero-Castano2009}.
Similar to these two molecules, CN is extremely resistant to the high UV irradiation from the central cluster.
Moreover, its chemistry is linked to that of HCN.
These species are shown to be widespread at the closest distances from Sgr~A$^*$.
The emission from H$_2$CO differs drastically from CN and it is mostly located in the outer regions of the CND tracing a $\sim 37''$ shell-like structure centered $\sim16''$ away from Sgr~A$^*$ where
H$_2$CO is observed to be even brighter than CN (with CN/H$_2$CO line ratios below 0.2). Mostly formed in solid phase, 
H$_2$CO emission in gas phase can be interpreted as the result of the expansion of Sgr~A East against the 50~\kms~ and 20~\kms~ GMCs east
and south of the CND, respectively.
Moreover, the $^{12}$C/$^{13}$C isotopic ratio, estimated from the $^{12}$CN/$^{13}$CN, supports the differentiation of the northeast arm, where the brightest H$_2$CO emission is observed.
While we estimate an opacity corrected carbon isotopic ratio in the range of $15-45$ throughout the CND, consistent with the idea of $^{13}$C enrichment with decreasing galactocentric distance due to nuclear processing \citep{Wilson1994,Wilson1999}, 
we find the lowest measured ratios ($<10$) towards the northeast arm.
The opacity derived from the hyperfine fit to the CN spectral profiles shows that CN is optically thick throughout most of the CND.
On the other hand, 
the densest ($n>10^6\rm cm^{-3}$) regions in the CND are located in the southern CND and the northeast arm as traced by HC$_3$N and SiO, although their chemical formation paths are different.
The southeast lobe and southern extension, where we find the largest SiO abundance and broader line widths, could be fueled and shocked by the molecular material from two streamers connecting the 20~\kms\, GMC,
where SiO abundance would be enhanced through grain erosion by low velocity shocks.
Similarly, the bright emission in the northeast lobe could be the consequence of the inflowing material from the 50~\kms~ streamer connecting  the GMC east of the CND.
The similar line widths measured in all species allow us to discard high velocity shocks as the origin of SiO.
The dense material in the northeast arm could be compressed due to the expansion of Sgr\,A East against the 50\,\kms\, GMC, where the brightest HC$_3$N emission as well as significant SiO emission are detected.
Even though HC$_3$N is efficiently formed through reactions involving CN, the emission of HC$_3$N is deficient close to Sgr~A$^*$ due to the high photodissociation rate of this molecule in UV exposed material.
The few regions in the inner CND where HC$_3$N emission is observed, will be tracing the densest cores ($A_v>1-2$) for this molecule to survive.
Finally, the carbon chain $\rm c-C_3H_2$ is observed to follow the emission of HC$_3$N, but with $\rm c-C_3H_2$/HC$_3$N at least an order of magnitude lower than in other Galactic PDRs. It is therefore unclear
why the emission of $\rm c-C_3H_2$ is so faint in the CND.

In the direction of Sgr~A$^*$ only CN is detected in absorption.
It shows two previously identified narrow absorptions at $-157$ and $-23$~\kms due to galactic arms along the line of sight.
A broad $94$~\kms\, line at $-32$~\kms\, is likely tracing the molecular gas closer to Sgr~A$^*$ absorbing the non-thermal central emission.
Moreover, we find hints of a possible wider ($>100$~\kms) absorption which might be tracing the closest molecular material to the central black hole.
However, observations of higher energy transitions of UV-resistant species are required to detect this molecular gas component in emission as shown by the HCN $4-3$ maps by \citet{Montero-Castano2009}.

We have studied the kinematics of all detected molecular clumps in the CND, where we considered kinematics of each individual velocity component at every selected position.
Our study shows most of these clumps are not associated with any rotating structure around Sgr~A$^*$ but with larger scale structures overlapping with the CND.
However, even if only $29\%$ of the molecular selected components appear clearly associated with rotating structures, they represent $70-75\%$ of the mass of the molecular gas.
Only three molecular structures show kinematic evidences of rotation around Sgr~A$^*$.
These rotating filaments, namely the southwest lobe, southern extension, and northeast lobe, are rotating in different planes, similar to what is derived from the study of the ionized arcs \citep{Zhao2009},
and are associated with the ionized mini-spiral structure observed in radio continuum and hydrogen recombination lines.

Moreover, we have shown how the various chemically differentiated structures are directly related to the different kinematical structures.
In this comparison we observe how CN is mostly concentrated in the dense molecular optically thick clumps rotating (and likely infalling)
filaments around Sgr~A$^*$ while the shell-like structure traced by H$_2$CO
is not orbiting around the central black hole, but shows velocities related to the overlapping GMCs.

This paper has shown the potential of detailed high resolution chemical observations to disentangle the complicated molecular structure surrounding Sgr~A$^*$,
as well as to establish the leading heating mechanisms affecting the different molecular structures in the region.
Similar to the low resolution work by \citet{Amo-Baladr'on2011}, chemical abundances from deep high resolution imaging of various molecular species as a function of position and distance from  Sgr~A$^*$ could potentially be used
to establish the three dimensional distribution of the molecular components in the Galactic center region.

\begin{acknowledgements}
J.M.-P. and S.M. have been supported MICINN through grants ESP2007-65812-C02-C01 and AYA2010-21697-C05-01, and AstroMadrid (CAM S2009/ESP-1496).
The authors would like to thank the referee, whose comments helped to significantly improve the paper.
\end{acknowledgements}

\bibliographystyle{aa}	
\bibliography{STH.bib}	

\Online





%


\begin{table*}[!p]
\begin{center}
\caption{CN and $^{13}$CN Gaussian fit parameters on selected positions \label{tab.CNvs13CN}}
\scriptsize
\begin{tabular}{c l    r@{$\pm$}l  r@{$\pm$}l  r@{$\pm$}l r@{$\pm$}l      r@{$\pm$}l r@{$\pm$}l r@{$\pm$}l     r@{$\pm$}l c}
\hline
\hline
Pos.      &  $\Delta \alpha,\Delta \delta$ \tfm{a}   &  \mc{8}{c}{CN}                                                                                                            & \mc{6}{c}{$^{13}$CN}                                                                      &  \mc{2}{c}{$^{12}$CN/$^{13}$CN \tfm{b}}  &  $^{12}$C/$^{13}$C \tfm{c} \\    
          &           ($''$,$''$)                    &  \mc{2}{c}{$\int S \delta v$}           & \mc{2}{c}{$v_{\rm LSR}$}   & \mc{2}{c}{$\Delta v$}  & \mc{2}{c}{$\tau$}\tfm{d}  & \mc{2}{c}{ $\int S \delta v$}      &  \mc{2}{c}{$v_{\rm LSR}$}   & \mc{2}{c}{$\Delta v$}  &  \mc{2}{c}{}                      &                       \\    
          &                                          &  \mc{2}{c}{(Jy \kms)}                   & \mc{2}{c}{\kms}            & \mc{2}{c}{\kms}        & \mc{2}{c}{}               & \mc{2}{c}{(Jy \kms)}               &  \mc{2}{c}{\kms}            & \mc{2}{c}{\kms}        &  \mc{2}{c}{}                      &                       \\    
\hline                                                                                                                                                                                                                                                                                                                                   
1         &   $(-26,-19) $                           &  200       &      30                    &    -47.7     &    1.2      &   59       &   3       &     3.5     &   0.8       &     9.3     &       1.8            &   -30       &   3           &    42       &   6      &    21 &    5                      &    26                 \\
2         &   $(-20,-22) $                           &  250       &      17                    &    -68.1     &    0.7      &   73.2     &   1.8     &     2.8     &   0.3       &    11       &       6              &   -46       &   6           &    61       &  16      &    23 &   12                      &    27                 \\
          &                                          &    8       &       2                    &     41.6     &    1.8      &   17       &   4       &     0.1     &   2.0       &     \mc{2}{c}{...}                 &     \mc{2}{c}{...}          &     \mc{2}{c}{...}     &    \mc{2}{c}{...}                 &                       \\
3         &   $(-20,-30) $                           &  120       &     110                    &   -116       &   10        &   40       &  30       &     0.5     &   0.1       &     5       &       2              &  -106       &   7           &    45       &  16      &    23 &   23                      &    23                 \\
          &                                          &   90       &      60                    &    -66       &   10        &   50       &  30       &     1.4     &   0.1       &     2.2     &       1.4            &   -39       &   0.0         &    30       &  13      &    40 &   37                      &    43                 \\
4         &   $(-20,-35) $                           &  198       &      15                    &   -120.3     &    0.4      &   46.1     &   1.4     &     1.0     &   0.3       &     8       &       2              &  -118       &   5           &    40       &   9      &    26 &    8                      &    28                 \\
5         &   $(-19,-42) $                           &  170       &      11                    &   -111.6     &    0.6      &   64.8     &   1.6     &     2.4     &   0.3       &     4.2     &       1.5            &   -89       &  14           &    60       &   0.0    &    40 &   15                      &    47                 \\
6         &   $(-2,-38)  $                           &  200       &     160                    &    -79       &   10        &   40       &  30       &     0.3     &   0.1       &     8.5     &       1.6            &   -72       &   5           &    40       &   0.0    &    24 &   20                      &    23                 \\
          &                                          &    8       &      15                    &     -9       &   10        &   20       &  30       &     0.1     &   0.1       &     \mc{2}{c}{...}                 &     \mc{2}{c}{...}          &     \mc{2}{c}{...}     &    \mc{2}{c}{...}                 &                       \\
7         &   $(1,-34)   $                           &  214       &       2                    &    -74.6     &    0.2      &   36.9     &   0.3     &     2.32    &   0.04      &    12       &       3              &   -72       &   4           &    52       &  11      &    19 &    5                      &    21                 \\
8         &   $(5,-25)   $                           &  181.8     &       1.8                  &    -60.92    &    0.04     &   18.14    &   0.15    &     1.72    &   0.02      &     7.6     &       0.7            &   -61.5     &   1.1         &    18       &   0.0    &    24 &    2                      &    26                 \\
9         &   $(15,-28)  $                           &   65       &       5                    &    -62.6     &    0.2      &   21.8     &   0.9     &     0.9     &   0.3       &     \mc{2}{c}{...}                 &     \mc{2}{c}{...}          &    \mc{2}{c}{...}      &    \mc{2}{c}{...}                 &                       \\
          &                                          &   12.0     &       1.3                  &     -3.4     &    0.6      &   16.8     &   1.4     &     0.1     &   0.2       &     \mc{2}{c}{...}                 &     \mc{2}{c}{...}          &    \mc{2}{c}{...}      &    \mc{2}{c}{...}                 &                       \\
          &                                          &   13       &       0.0                  &     52       &    2        &   24       &   4       &     4       &   2         &     3       &       3              &    42       &  10           &    22       &  14      &     4 &    4                      &     6                 \\
10        &   $(-23,-5)  $                           &   57       &       7                    &     19.3     &    0.5      &   31.0     &   1.7     &     0.8     &   0.5       &     3       &       3              &    36       &  19           &    50       &  30      &    21 &   22                      &    22                 \\
          &                                          &   23       &       5                    &     71.6     &    1.0      &   26   3   &   2       &     0.4     &   0.7       &     \mc{2}{c}{...}                 &     \mc{2}{c}{...}          &    \mc{2}{c}{...}      &    \mc{2}{c}{...}                 &                       \\
11        &   $(-24,5)   $                           &   17       &       3                    &     -4.6     &    0.8      &   20       &   2       &     0.1     &   0.7       &     \mc{2}{c}{...}                 &     \mc{2}{c}{...}          &    \mc{2}{c}{...}      &    \mc{2}{c}{...}                 &                       \\
          &                                          &   20       &       3                    &     68.3     &    0.2      &   17       &   2       &     0.1     &   0.3       &     \mc{2}{c}{...}                 &     \mc{2}{c}{...}          &    \mc{2}{c}{...}      &    \mc{2}{c}{...}                 &                       \\
12        &   $(15,-2)   $                           &   26       &       7                    &    -36.1     &    1.5      &   26       &   3       &     4.8     &   1.7       &     \mc{2}{c}{...}                 &     \mc{2}{c}{...}          &    \mc{2}{c}{...}      &    \mc{2}{c}{...}                 &                       \\
          &                                          &   32       &       3                    &      4.3     &    0.4      &   16.65    &   0.15    &     1.9     &   0.5       &     \mc{2}{c}{...}                 &     \mc{2}{c}{...}          &    \mc{2}{c}{...}      &    \mc{2}{c}{...}                 &                       \\
13        &   $(9,-17)   $                           &   60       &       9                    &    -64.3     &    1.7      &   45       &   3       &     2.3     &   0.6       &     \mc{2}{c}{...}                 &     \mc{2}{c}{...}          &    \mc{2}{c}{...}      &    \mc{2}{c}{...}                 &                       \\
          &                                          &   69       &       5                    &    -39.1     &    0.2      &   16.65    &   0.05    &     2.4     &   0.3       &     6.2     &       1.7            &   -41       &   2           &    25       &   5      &    11 &    3                      &    13                 \\
14        &   $(25,-7)   $                           &   37       &       9                    &    -42.3     &    1.2      &   30       &   3       &     0.9     &   0.9       &     \mc{2}{c}{...}                 &     \mc{2}{c}{...}          &    \mc{2}{c}{...}      &    \mc{2}{c}{...}                 &                       \\
          &                                          &    6.8     &       1.8                  &    121.1     &    0.7      &   32.1     &   1.1     &     0.10    &   0.03      &     \mc{2}{c}{...}                 &     \mc{2}{c}{...}          &    \mc{2}{c}{...}      &    \mc{2}{c}{...}                 &                       \\
15        &   $(12,13)   $                           &   57       &      10                    &     11.6     &    0.9      &   24.5     &   1.8     &     7.3     &   1.6       &     \mc{2}{c}{...}                 &     \mc{2}{c}{...}          &    \mc{2}{c}{...}      &    \mc{2}{c}{...}                 &                       \\
16        &   $(21,11)   $                           &   20       &      18                    &    -12       &   10        &   36       &  30       &     0.1     &   0.1       &     2.7     &       0.6            &    33       &   3           &    19.5     &   0.6    &     7 &    7                      &     9                 \\
          &                                          &   26       &      51                    &     36       &   10        &   17       &  30       &     0.1     &   0.1       &     \mc{2}{c}{...}                 &     \mc{2}{c}{...}          &    \mc{2}{c}{...}      &    \mc{2}{c}{...}                 &                       \\
17        &   $(-8,24)   $                           &   69       &       5                    &     39.2     &    0.2      &   25.3     &   0.6     &     1.1     &   0.2       &     3.7     &       0.9            &    39       &   0.0         &    25       &   0.0    &    19 &    5                      &    20                 \\
18        &   $(1,5)     $                           &    3.8     &       1.0                  &    -77       &    2        &   16.6     &   1.6     &     0.1     &   0.8       &     \mc{2}{c}{...}                 &     \mc{2}{c}{...}          &    \mc{2}{c}{...}      &    \mc{2}{c}{...}                 &                       \\
          &                                          &   35       &       6                    &     45.8     &    0.6      &   19.0     &   1.3     &     2.4     &   0.8       &     \mc{2}{c}{...}                 &     \mc{2}{c}{...}          &    \mc{2}{c}{...}      &    \mc{2}{c}{...}                 &                       \\
19        &   $(0,0)     $                           &   \mc{2}{c}{...}                        &     \mc{2}{c}{...}         &   \mc{2}{c}{...}       &    \mc{2}{c}{...}         &    \mc{2}{c}{...}                  &     \mc{2}{c}{...}          &    \mc{2}{c}{...}      &    \mc{2}{c}{...}                 &                       \\
20        &   $(11,27)   $                           &   10       &       3                    &    -68       &    4        &   31       &   6       &    10       &   4         &    \mc{2}{c}{...}                  &     \mc{2}{c}{...}          &    \mc{2}{c}{...}      &    \mc{2}{c}{...}                 &                       \\
          &                                          &   54       &       2                    &     60.3     &    0.4      &   22.3     &   0.7     &     0.10    &   0.05      &     \mc{2}{c}{...}                 &     \mc{2}{c}{...}          &    \mc{2}{c}{...}      &    \mc{2}{c}{...}                 &                       \\
          &                                          &   17       &       0.0                  &     89.2     &    1.3      &   24       &   2       &     0.1     &   0.1       &     \mc{2}{c}{...}                 &     \mc{2}{c}{...}          &    \mc{2}{c}{...}      &    \mc{2}{c}{...}                 &                       \\
21        &   $(12,41)   $                           &  101       &      16                    &    113.6     &    0.7      &   28       &   2       &     2.6     &   0.7       &     4       &       2              &   116       &   4           &    21       &  10      &    26 &   17                      &    31                 \\
22        &   $(25,39)   $                           &  151       &       8                    &     86.9     &    0.2      &   29.0     &   0.7     &     2.0     &   0.2       &     5.4     &       1.0            &    90       &   4           &    29       &   0.0    &    28 &    5                      &    32                 \\
23        &   $(28,34)   $                           &  160       &       8                    &     80.0     &    0.2      &   39.6     &   0.8     &     0.78    &   0.18      &     6       &       2              &    76       &   6           &    37       &   8      &    27 &    9                      &    28                 \\
24        &   $(-3,40)   $                           &    8       &       10                   &    -84       &   10        &   30       &  30       &     2.3     &   0.1       &     \mc{2}{c}{...}                 &     \mc{2}{c}{...}          &    \mc{2}{c}{...}      &    \mc{2}{c}{...}                 &                       \\
          &                                          &   24       &      30                    &     31       &   10        &   30       &  30       &     0.3     &   0.1       &     4.9     &       1.2            &    31       &   0.0         &    25       &   0.0    &     5 &    7                      &     5                 \\
          &                                          &   25       &       0.0                  &     70       &   10        &   30       &  30       &     0.1     &   0.1       &     \mc{2}{c}{...}                 &     \mc{2}{c}{...}          &    \mc{2}{c}{...}      &    \mc{2}{c}{...}                 &                       \\
25        &   $(-15,40)  $                           &   40       &      10                    &     69.9     &    1.4      &   31   8   &   4       &     4.5     &   1.5       &     3.7     &       1.2            &    70       &   0.0         &    31       &   0.0    &    11 &    5                      &    15                 \\
26        &   $(-21,33)  $                           &   42       &      10                    &     38.7     &    0.8      &   21       &   2       &     4.0     &   1.2       &     \mc{2}{c}{...}                 &     \mc{2}{c}{...}          &    \mc{2}{c}{...}      &    \mc{2}{c}{...}                 &                       \\
27        &   $(-29,31)  $                           &    7.0     &       0.7                  &    -79.5     &    0.9      &   16.6     &   0.7     &     0.1     &   0.5       &     \mc{2}{c}{...}                 &     \mc{2}{c}{...}          &    \mc{2}{c}{...}      &    \mc{2}{c}{...}                 &                       \\
          &                                          &   55.3     &       1.8                  &     44.4     &    0.2      &   16.65    &   0.07    &     2.1     &   0.2       &     \mc{2}{c}{...}                 &     \mc{2}{c}{...}          &    \mc{2}{c}{...}      &    \mc{2}{c}{...}                 &                       \\
28        &   $(-3,45)   $                           &   13       &       2                    &     80.7     &    1.1      &   20       &   3       &     0.1     &   2         &     \mc{2}{c}{...}                 &     \mc{2}{c}{...}          &    \mc{2}{c}{...}      &    \mc{2}{c}{...}                 &                       \\
29        &   $(7,46)    $                           &    6       &       3                    &    -37       &    3        &   16       &   5       &     0.1     &   0.0       &     \mc{2}{c}{...}                 &     \mc{2}{c}{...}          &    \mc{2}{c}{...}      &    \mc{2}{c}{...}                 &                       \\
          &                                          &   21       &       4                    &    108.1     &    1.1      &   17       &   2       &     0.1     &   0.0       &     \mc{2}{c}{...}                 &     \mc{2}{c}{...}          &    \mc{2}{c}{...}      &    \mc{2}{c}{...}                 &                       \\
30        &   $(48,24)   $                           &   29.8     &       1.7                  &     49.1     &    0.4      &   16.6     &   0.6     &     0.1     &   0.0       &     \mc{2}{c}{...}                 &     \mc{2}{c}{...}          &    \mc{2}{c}{...}      &    \mc{2}{c}{...}                 &                       \\
31        &   $(46,20)   $                           &   15.9     &       1.1                  &     48.5     &    0.7      &   16.66    &   0.02    &     0.2     &   0.0       &     3       &       2              &    48       &   0.0         &    17       &   0.0    &     5 &    3                      &     5                 \\
32        &   $(50,9)    $                           &   14       &       2                    &    -56.4     &    1.7      &   38       &   4       &     0.42    &   0.15      &     \mc{2}{c}{...}                 &     \mc{2}{c}{...}          &    \mc{2}{c}{...}      &    \mc{2}{c}{...}                 &                       \\
          &                                          &   17.5     &       0.7                  &     38.8     &    0.4      &   16.6     &   0.1     &     0.145   &   0.002     &     5.1     &       1.2            &    35.0     &   1.8         &    17       &   2      &     3 &   0.8                     &     3                 \\
33        &   $(51,1)    $                           &    6.6     &       1.3                  &    -60.1     &    1.3      &   16       &   2       &     0.1     &   0.5       &     \mc{2}{c}{...}                 &     \mc{2}{c}{...}          &    \mc{2}{c}{...}      &   \mc{2}{c}{...}                  &                       \\
          &                                          &   11       &       2                    &      4.1     &    0.9      &   16.6     &   0.6     &     1.2     &   1.1       &     \mc{2}{c}{...}                 &     \mc{2}{c}{...}          &    \mc{2}{c}{...}      &   \mc{2}{c}{...}                  &                       \\
          &                                          &   18       &       0                    &     45.9     &    0.5      &   16.7     &   1.3     &     0.4     &   0.5       &     6       &       2              &    35.9     &   1.6         &    17       &   4      &     3 &   0.9                     &     3                 \\
34        &   $(46,-4)   $                           &   10       &       6                    &    -51       &    4.0      &   33       &   7       &     2       &   3         &     \mc{2}{c}{...}                 &     \mc{2}{c}{...}          &    \mc{2}{c}{...}      &   \mc{2}{c}{...}                  &                       \\
          &                                          &   14.2     &       0.8                  &      3.3     &    0.6      &   16.6     &   0.2     &     0.1     &   0.2       &     \mc{2}{c}{...}                 &     \mc{2}{c}{...}          &    \mc{2}{c}{...}      &   \mc{2}{c}{...}                  &                       \\
35        &   $(39,-2)   $                           &   15.6     &       1.5                  &      2.8     &    0.8      &   16.6     &   0.2     &     0.1     &   0.1       &     \mc{2}{c}{...}                 &     \mc{2}{c}{...}          &    \mc{2}{c}{...}      &   \mc{2}{c}{...}                  &                       \\
36        &   $(18,-34)  $                           &   \mc{2}{c}{...}                        &     \mc{2}{c}{...}         &   \mc{2}{c}{...}       &     \mc{2}{c}{...}        &     \mc{2}{c}{...}                 &     \mc{2}{c}{...}          &    \mc{2}{c}{...}      &   \mc{2}{c}{...}                  &                       \\   
37        &   $(-6,-36)  $                           &   65       &       8                    &    -79.1     &    0.6      &   33.4     &   1.7     &     0.4     &   0.4       &     4       &       2              &   -85       &   4           &    21       &  12      &    18 &   13                      &    18                 \\                                                                                                      
          &                                          &    6.4     &       1.3                  &     -6.0     &    1.7      &   16.6     &   1.5     &     0.1     &   0.5       &     \mc{2}{c}{...}                 &     \mc{2}{c}{...}          &    \mc{2}{c}{...}      &   \mc{2}{c}{...}                  &                       \\
38        &   $(-14,9)   $                           &   33       &      60                    &     27       &   10        &   20       &  30       &     4.7     &   0.1       &     \mc{2}{c}{...}                 &     \mc{2}{c}{...}          &    \mc{2}{c}{...}      &   \mc{2}{c}{...}                  &                       \\
          &                                          &    7       &      11                    &     86       &   10        &   20       &  30       &     0.1     &   0.1       &     \mc{2}{c}{...}                 &     \mc{2}{c}{...}          &    \mc{2}{c}{...}      &   \mc{2}{c}{...}                  &                       \\
39        &   $(-19,21)  $                           &   11.8     &       1.0                  &     67.6     &    0.7      &   16.6     &   0.7     &     0.1     &   0.4       &     \mc{2}{c}{...}                 &     \mc{2}{c}{...}          &    \mc{2}{c}{...}      &   \mc{2}{c}{...}                  &                       \\
\hline                                                                       
\end{tabular}
\tablefoot{
\tft{a}{Offset positions are relative to the nominal center of the mosaic $\alpha_{\rm J2000}=17^{\rm h}45^{\rm m}40^{\rm s},\,\delta_{\rm J2000}=-29^\circ00'28''$.}
\tft{b}{Ratio of total integrated intensity of the CN and $^{13}$CN spectral features.}
\tft{c}{Carbon isotopic ratio estimated from the opacity corrected CN/$^{13}$CN integrated intensity ratio.}
\tft{d}{Peak optical depth of the main hyperfine component.}
}
\end{center}
\end{table*}

\newpage

\begin{sidewaystable*}[!p]
\begin{center}
\caption{H$_2$CO, SiO, c-C$_3$H$_2$ and HC$_3$N Gaussian fit parameters on selected positions \label{tab.GaussFits}}
\scriptsize
\begin{tabular}{c l   r@{$\pm$}l r@{$\pm$}l r@{$\pm$}l    r@{$\pm$}l r@{$\pm$}l r@{$\pm$}l   r@{$\pm$}l r@{$\pm$}l r@{$\pm$}l   r@{$\pm$}l r@{$\pm$}l r@{$\pm$}l} 
\hline
\hline
Pos.      &  $\Delta \alpha,\Delta \delta$  \tfm{a}  &  \mc{6}{c}{H$_2$CO}                                                                  &  \mc{6}{c}{SiO}                                                                     &  \mc{6}{c}{c-C$_3$H$_2$}                                                                        &  \mc{6}{c}{HC$_3$N}                                  \\
          &           ($''$,$''$)                    &  \mc{2}{c}{$\int S \delta v$}  &  \mc{2}{c}{$v_{\rm LSR}$}  & \mc{2}{c}{$\Delta v$}  &  \mc{2}{c}{$\int S \delta v$}  &  \mc{2}{c}{$v_{\rm LSR}$}  & \mc{2}{c}{$\Delta v$} &   \mc{2}{c}{$\int S \delta v$}  &   \mc{2}{c}{$v_{\rm LSR}$}    &  \mc{2}{c}{$\Delta v$}        &   \mc{2}{c}{$\int S \delta v$}  &   \mc{2}{c}{$v_{\rm LSR}$}    &  \mc{2}{c}{$\Delta v$ }         \\
          &                                          &  \mc{2}{c}{(Jy \kms)}          &  \mc{2}{c}{\kms}           & \mc{2}{c}{\kms}        &  \mc{2}{c}{(Jy \kms)}          &  \mc{2}{c}{\kms}           & \mc{2}{c}{\kms}       &   \mc{2}{c}{(Jy \kms)}          &   \mc{2}{c}{\kms}             &  \mc{2}{c}{\kms}              &   \mc{2}{c}{(Jy \kms)}          &   \mc{2}{c}{\kms}             &  \mc{2}{c}{\kms}                \\
\hline                                                                                                                                                                                                                                                               
1         &   $(-26,-19) $                           & 7.2      &   0.8               &  -38     &  3              &  44     &  5           &  29.7    &  0.8                &  -43.8   &  0.8            & 55.3   &  1.7         &  2.8    &   0.6                 &  -45     &  6                 &  46      & 10                 &   5.0    &  0.7                 & -47     &  5                  &  50      &  0.0                 \\                                 
2         &   $(-20,-22) $                           & 1.1      &   0.4               &  -97     &  0.0            &  26     &  0.0         &  4.0     &  0.8                &  -97     &  3              & 26     &  5           &  1.9    &   0.7                 &  -56     & 13                 &  68      & 23                 &   3.8    &  0.9                 & -65     &  9                  &  74      & 18                   \\ 
          &                                          & 6.2      &   0.7               &  -38     &  3              &  47     &  6           &  19.5    &  1.0                &  -39.1   &  1.3            & 49     &  3           &     \mc{2}{c}{...}              &      \mc{2}{c}{...}           &      \mc{2}{c}{...}           &      \mc{2}{c}{...}             &     \mc{2}{c}{...}            &      \mc{2}{c}{...}             \\      
          &                                          & 1.5      &   0.4               &   43     &  2              &  14     &  4           &  \mc{2}{c}{...}                &      \mc{2}{c}{...}        &    \mc{2}{c}{...}     &     \mc{2}{c}{...}              &      \mc{2}{c}{...}           &      \mc{2}{c}{...}           &      \mc{2}{c}{...}             &     \mc{2}{c}{...}            &      \mc{2}{c}{...}             \\
3         &   $(-20,-30) $                           & 1.8      &   0.6               & -109     &  5              &  31     & 10           &  9.1     &  0.7                & -111.4   &  1.9            & 47     &  4           &  1.3    &   0.3                 & -108     &  6                 &  26      &  7                 &      \mc{2}{c}{...}             &     \mc{2}{c}{...}            &      \mc{2}{c}{...}             \\                                 
          &                                          & 1.8      &   0.6               &  -36     &  6              &  39     &  0.0         &  6.6     &  0.6                &  -40     &  2              & 40     &  4           &     \mc{2}{c}{...}              &      \mc{2}{c}{...}           &      \mc{2}{c}{...}           &      \mc{2}{c}{...}             &     \mc{2}{c}{...}            &      \mc{2}{c}{...}             \\  
4         &   $(-20,-35) $                           & 4.2      &   0.8               & -123     &  3              &  34     &  8           &  19.4    &  0.8                & -123.1   &  0.7            & 36.6   &  1.7         &  1.2    &   0.7                 &  -95     & 18                 &  52      & 30                 &   2.7    &  0.6                 &-126     &  3                  &  26      &  7                   \\                                 
5         &   $(-19,-42) $                           & 3.8      &   1.0               & -100     &  8              &  59     & 18           &  7.2     &  1.2                & -102.2   &  5              & 60     & 10           &  1.5    &   0.8                 &  -85     & 13                 &  60      &  0.0               &      \mc{2}{c}{...}             &     \mc{2}{c}{...}            &      \mc{2}{c}{...}             \\                                 
6         &   $(-2,-38)  $                           & 4.2      &   0.8               &  -71     &  4              &  37     &  7           &  14.0    &  1.0                &  -77.3   &  1.1            & 34     &  3           &  2.8    &   0.6                 &  -77     &  3                 &  30      &  4                 &   2.2    &  0.6                 & -70     &  3                  &  22      &  8                   \\ 
          &                                          & 3.8      &   0.5               &  -10.1   &  1.2            &  14     &  3           &  \mc{2}{c}{...}                &      \mc{2}{c}{...}        &    \mc{2}{c}{...}     &     \mc{2}{c}{...}              &      \mc{2}{c}{...}           &      \mc{2}{c}{...}           &      \mc{2}{c}{...}             &     \mc{2}{c}{...}            &      \mc{2}{c}{...}             \\                      
7         &   $(1,-34)   $                           & 4.8      &   0.7               &  -71.4   &  2              &  33     &  5           &  18.2    &  0.8                &  -75.7   &  0.8            & 37     &  2           &  2.5    &   0.7                 &  -79     &  3                 &  24      & 10                 &   1.5    &  0.5                 & -65     &  6                  &  28      &  8                   \\                                 
8         &   $(5,-25)   $                           & 4.3      &   0.5               &  -63.0   &  1.1            &  20     &  2           &  6.7     &  0.6                &  -54.4   &  1.2            & 28     &  3           &  2.0    &   0.4                 &  -59     &  1                 &  15      &  3                 &   1.9    &  0.8                 & -58     &  4                  &  25      & 17                   \\                                 
9         &   $(15,-28)  $                           &  \mc{2}{c}{...}                &   \mc{2}{c}{...}           &  \mc{2}{c}{...}        &  1.5     &  0.5                &  -54.0   &  1.4            & 10     &187           &     \mc{2}{c}{...}              &      \mc{2}{c}{...}           &      \mc{2}{c}{...}           &      \mc{2}{c}{...}             &     \mc{2}{c}{...}            &      \mc{2}{c}{...}             \\
          &                                          & 13.3     &   0.6               &    6.8   &  0.6            &  25.8   &  1.3         &  2.4     &  0.6                &    6.9   &  3.0            & 20     &  6           &  0.7    &   0.3                 &   -3     &  2                 &  10      & 18                 &      \mc{2}{c}{...}             &     \mc{2}{c}{...}            &      \mc{2}{c}{...}             \\
10        &   $(-23,-5)  $                           & 1.1      &   0.4               &   19     &  5              &  11     &112           &  3.4     &  0.5                &   12.6   &  1.2            & 18     &  4           &     \mc{2}{c}{...}              &      \mc{2}{c}{...}           &      \mc{2}{c}{...}           &      \mc{2}{c}{...}             &     \mc{2}{c}{...}            &      \mc{2}{c}{...}             \\ 
          &                                          & 2.6      &   0.5               &   72     &  2              &  24     &  4           &  1.0     &  0.4                &   63     &  2              & 10     & 21           &     \mc{2}{c}{...}              &      \mc{2}{c}{...}           &      \mc{2}{c}{...}           &      \mc{2}{c}{...}             &     \mc{2}{c}{...}            &      \mc{2}{c}{...}             \\
11        &   $(-24,5)   $                           & 3.2      &   0.5               &   68.6   &  1.6            &  20     &  4           &  2.0     &  0.5                &   64     &  3              & 24     &  7           &     \mc{2}{c}{...}              &      \mc{2}{c}{...}           &      \mc{2}{c}{...}           &      \mc{2}{c}{...}             &     \mc{2}{c}{...}            &      \mc{2}{c}{...}             \\                                  
12        &   $(15,-2)   $                           &  \mc{2}{c}{...}                &   \mc{2}{c}{...}           &  \mc{2}{c}{...}        &  \mc{2}{c}{...}                &      \mc{2}{c}{...}        &    \mc{2}{c}{...}     &     \mc{2}{c}{...}              &      \mc{2}{c}{...}           &      \mc{2}{c}{...}           &   1.4    &  0.4                 &  15.6   &  1.6                &  17      &  7                   \\                                  
13        &   $(9,-17)   $                           & 3.6      &   0.7               &  -41     &  3              &  27     &  6           &  3.6     &  0.4                &  -36.5   &  0.9            & 17     &  2           &  0.6    &   0.2                 &  -30     &  6                 &  10      & 62                 &   2.9    &  0.8                 & -67     & 10                  &  72      & 22                   \\                                  
14        &   $(25,-7)   $                           &  \mc{2}{c}{...}                &   \mc{2}{c}{...}           &  \mc{2}{c}{...}        &  1.3     &  0.8                &  -43     & 14              & 57     & 56           &     \mc{2}{c}{...}              &      \mc{2}{c}{...}           &      \mc{2}{c}{...}           &   1.7    &  0.7                 & -52     & 13                  &  58      & 30                   \\                                  
15        &   $(12,13)   $                           & 1.3      &   0.3               &   17.7   &  0.9            &  10     & 54           &  2.2     &  0.6                &   24     &  5              & 31     &  7           &  1.0    &   0.3                 &   14     &  1.2               &  10      & 31                 &      \mc{2}{c}{...}             &     \mc{2}{c}{...}            &      \mc{2}{c}{...}             \\                                  
16        &   $(21,11)   $                           &0.9       &   0.4               &  -27     &  3              &  14     &  5           &  2.8     &  0.5                &  -17     &  4              & 37     &  7           &     \mc{2}{c}{...}              &      \mc{2}{c}{...}           &      \mc{2}{c}{...}           &      \mc{2}{c}{...}             &     \mc{2}{c}{...}            &      \mc{2}{c}{...}             \\                                  
          &                                          &0.8       &   0.4               &   41     &  5              &  19     & 11           & 0.8      &  0.3                &   34.8   &  1.2            & 10     & 18           &     \mc{2}{c}{...}              &      \mc{2}{c}{...}           &      \mc{2}{c}{...}           &      \mc{2}{c}{...}             &     \mc{2}{c}{...}            &      \mc{2}{c}{...}             \\
17        &   $(-8,24)   $                           & 2.7      &   0.6               &   24     &  2              &  21     &  5           &  4.2     &  0.6                &   29     &  2              & 32     &  6           &     \mc{2}{c}{...}              &      \mc{2}{c}{...}           &      \mc{2}{c}{...}           &   2.2    &  0.6                 &  21     &  4                  &  28      &  7                   \\                                  
18        &   $(1,5)     $                           &  \mc{2}{c}{...}                &   \mc{2}{c}{...}           &  \mc{2}{c}{...}        &  \mc{2}{c}{...}                &      \mc{2}{c}{...}        &    \mc{2}{c}{...}     &     \mc{2}{c}{...}              &      \mc{2}{c}{...}           &      \mc{2}{c}{...}           &      \mc{2}{c}{...}             &     \mc{2}{c}{...}            &      \mc{2}{c}{...}             \\                                  
19        &   $(0,0)     $                           &  \mc{2}{c}{...}                &   \mc{2}{c}{...}           &  \mc{2}{c}{...}        &  \mc{2}{c}{...}                &      \mc{2}{c}{...}        &    \mc{2}{c}{...}     &     \mc{2}{c}{...}              &      \mc{2}{c}{...}           &      \mc{2}{c}{...}           &      \mc{2}{c}{...}             &     \mc{2}{c}{...}            &      \mc{2}{c}{...}             \\
20        &   $(11,27)   $                           & 1.5      &   0.4               &  -12     &  2              &  18     &  6           &  \mc{2}{c}{...}                &      \mc{2}{c}{...}        &    \mc{2}{c}{...}     &     \mc{2}{c}{...}              &      \mc{2}{c}{...}           &      \mc{2}{c}{...}           &      \mc{2}{c}{...}             &     \mc{2}{c}{...}            &      \mc{2}{c}{...}             \\
          &                                          & 1.7      &   0.4               &   56     &  4              &  25     &  9           &  1.0     &  0.4                &   66     &  3              & 15     & 11           &  1.6    &   0.4                 &   70     &  6                 &  41      & 11                 &  0.7     &  0.4                 &  60     &  5                  &  19      & 14                   \\
21        &   $(12,41)   $                           &  \mc{2}{c}{...}                &   \mc{2}{c}{...}           &  \mc{2}{c}{...}        &  1.5     &  0.5                &  121     &  2              & 10     &  9           &     \mc{2}{c}{...}              &      \mc{2}{c}{...}           &      \mc{2}{c}{...}           &      \mc{2}{c}{...}             &     \mc{2}{c}{...}            &      \mc{2}{c}{...}             \\                                  
22        &   $(25,39)   $                           & 3.9      &   0.6               &   90.0   &  1.8            &  23     &  3           &  6.4     &  0.7                &   88.3   &  1.4            & 26     &  3           &  0.9    &   0.5                 &   92     &  5                 &  16      &  9                 &   1.9    &  0.6                 &  86     &  5                  &  30      & 10                   \\                                  
23        &   $(28,34)   $                           & 1.8      &   0.4               &   29.5   &  1.1            &  10     &  5           &  1.0     &  0.4                &   31     &  3              & 10     &  3           &     \mc{2}{c}{...}              &      \mc{2}{c}{...}           &      \mc{2}{c}{...}           &      \mc{2}{c}{...}             &     \mc{2}{c}{...}            &      \mc{2}{c}{...}             \\ 
          &                                          & 3.3      &   0.7               &   74.0   &  3              &  31     &  6           &  4.1     &  0.7                &   79     &  3              & 31     &  0.0         &  1.4    &   0.6                 &   77     &  6                 &  25      &  9                 &      \mc{2}{c}{...}             &     \mc{2}{c}{...}            &      \mc{2}{c}{...}             \\
24        &   $(-3,40)   $                           & 4.2      &   1.0               &   13     &  3              &  26     &  7           &  4.0     &  0.7                &   10     &  3              & 27     &  6           &     \mc{2}{c}{...}              &      \mc{2}{c}{...}           &      \mc{2}{c}{...}           &   2.0    &  1.0                 &  45     &  6                  &  28      & 20                   \\                                  
25        &   $(-15,40)  $                           & 4.0      &   0.6               &   75     &  2              &  25     &  3           &  2.1     &  0.7                &   76     &  4              & 23     &  8           &     \mc{2}{c}{...}              &      \mc{2}{c}{...}           &      \mc{2}{c}{...}           &      \mc{2}{c}{...}             &     \mc{2}{c}{...}            &      \mc{2}{c}{...}             \\                                  
26        &   $(-21,33)  $                           &  \mc{2}{c}{...}                &   \mc{2}{c}{...}           &  \mc{2}{c}{...}        &  \mc{2}{c}{...}                &      \mc{2}{c}{...}        &    \mc{2}{c}{...}     &     \mc{2}{c}{...}              &      \mc{2}{c}{...}           &      \mc{2}{c}{...}           &      \mc{2}{c}{...}             &     \mc{2}{c}{...}            &      \mc{2}{c}{...}             \\                                  
27        &   $(-29,31)  $                           & 1.2      &   0.4               &   47.2   &  1.6            &  10     & 53           &  \mc{2}{c}{...}                &      \mc{2}{c}{...}        &    \mc{2}{c}{...}     &  2.2    &   0.6                 &   47     &  2                 &  19      &  6                 &      \mc{2}{c}{...}             &     \mc{2}{c}{...}            &      \mc{2}{c}{...}             \\                                  
28        &   $(-3,45)   $                           & 3.2      &   1.0               &   -6     &  3              &  18     &  8           &  \mc{2}{c}{...}                &      \mc{2}{c}{...}        &    \mc{2}{c}{...}     &     \mc{2}{c}{...}              &      \mc{2}{c}{...}           &      \mc{2}{c}{...}           &      \mc{2}{c}{...}             &     \mc{2}{c}{...}            &      \mc{2}{c}{...}             \\
29        &   $(7,46)    $                           & 4.3      &   0.9               &  -26     &  3              &  27     &  7           &  2.2     &  0.8                &  -29     &  6              & 29     & 14           &     \mc{2}{c}{...}              &      \mc{2}{c}{...}           &      \mc{2}{c}{...}           &      \mc{2}{c}{...}             &     \mc{2}{c}{...}            &      \mc{2}{c}{...}             \\
30        &   $(48,24)   $                           & 11.2     &   1.7               &   44     &  2              &  26     &  4           &  12.1    &  0.9                &   42     &  1              & 30     &  2           &  3.4    &   0.8                 &   39     &  3                 &  23      &  6                 &   5.5    &  0.8                 &  41.4   &  1.7                &  24      &  4                   \\
31        &   $(46,20)   $                           & 8.0      &   0.5               &   43.0   &  0.7            &  19.2   &  1.3         &  10.6    &  0.8                &   44.3   &  0.9            & 24     &  2           &  4.6    &   0.8                 &   41     &  2                 &  24      &  6                 &   3.5    &  0.5                 &  42.9   &  1.4                &  19      &  3                   \\
32        &   $(50,9)    $                           & 10.6     &   1.5               &   35.6   &  1.2            &  18     &  3           &  4.0     &  0.6                &   36.3   &  1.2            & 15     &  3           &  3.6    &   0.6                 &   37.3   &  1.3               &  14      &  3                 &   2.6    &  0.4                 &  34.3   &  0.6                &  10      & 10                   \\    
33        &   $(51,1)    $                           & 8.8      &   1.7               &   33.5   &  1.9            &  20     &  4           &  5.3     &  0.7                &   36.8   &  1.5            & 23     &  4           &  4.7    &   0.9                 &   32.5   &  1.7               &  18      &  4                 &   5.8    &  0.8                 &  34.3   &  1.8                &  27      &  5                   \\
34        &   $(46,-4)   $                           & 6.4      &   0.6               &    4.6   &  0.8            &  10     &  2           &  1.9     &  0.6                &   10     &  3              & 20     &  6           &  1.8    &   0.4                 &  -52     &  2                 &  19      &  4                 &      \mc{2}{c}{...}             &     \mc{2}{c}{...}            &      \mc{2}{c}{...}             \\
35        &   $(39,-2)   $                           & 5.6      &   0.4               &    3.1   &  1.1            &  10     &  1.1         &  \mc{2}{c}{...}                &      \mc{2}{c}{...}        &    \mc{2}{c}{...}     &     \mc{2}{c}{...}              &      \mc{2}{c}{...}           &      \mc{2}{c}{...}           &      \mc{2}{c}{...}             &     \mc{2}{c}{...}            &      \mc{2}{c}{...}             \\
36        &   $(18,-34)  $                           & 5.3      &   0.8               &   -8.6   &  1.4            &  18     &  4           &  \mc{2}{c}{...}                &      \mc{2}{c}{...}        &    \mc{2}{c}{...}     &     \mc{2}{c}{...}              &      \mc{2}{c}{...}           &      \mc{2}{c}{...}           &      \mc{2}{c}{...}             &     \mc{2}{c}{...}            &      \mc{2}{c}{...}             \\
          &                                          & 6.9      &   0.5               &   11     &  0.0            &  12     &  0.0         &  3.5     &  0.6                &   11.1   &  1.5            & 12     &  2           &     \mc{2}{c}{...}              &      \mc{2}{c}{...}           &      \mc{2}{c}{...}           &      \mc{2}{c}{...}             &     \mc{2}{c}{...}            &      \mc{2}{c}{...}             \\
37        &   $(-6,-36)  $                           &  \mc{2}{c}{...}                &    \mc{2}{c}{...}          &  \mc{2}{c}{...}        &  5.7     &  0.6                &  -79.2   &  1.4            & 30     &  4           &  2.6    &   0.5                 &  -94     &  3                 &  30      &  0.0               &      \mc{2}{c}{...}             &     \mc{2}{c}{...}            &      \mc{2}{c}{...}             \\
          &                                          & 4.9      &   0.5               &   -6.0   &  0.9            &  14.4   &  1.5         &  2.3     &  0.6                &   12     &  4              & 32     &  8           &     \mc{2}{c}{...}              &      \mc{2}{c}{...}           &      \mc{2}{c}{...}           &      \mc{2}{c}{...}             &     \mc{2}{c}{...}            &      \mc{2}{c}{...}             \\
38        &   $(-14,9)   $                           & 5.3      &   0.8               &   31.4   &  1.7            &  23     &  4           &  3.4     &  0.5                &   34.9   &  1.6            & 24     &  4           &  2.4    &   0.7                 &   26     &  9                 &  64      & 23                 &   2.0    &  0.4                 &  32     &  2                  &  19      &  4                   \\
39        &   $(-19,21)  $                           & 4.8      &   0.9               &   59     &  3              &  27     &  7           &  1.4     &  0.5                &   56     &  3              & 14     & 12           &     \mc{2}{c}{...}              &      \mc{2}{c}{...}           &      \mc{2}{c}{...}           &   1.2    &  0.5                 &  60     &  3                  &  16      &  5                   \\
\hline                                           
\end{tabular}                                    
\tablefoot{
\tft{a}{Offset positions are relative to the nominal center of the mosaic $\alpha_{\rm J2000}=17^{\rm h}45^{\rm m}40^{\rm s},\,\delta_{\rm J2000}=-29^\circ00'28''$.}
}
\end{center}                                     
\,
\end{sidewaystable*}
                                                 
                                               
                                                 
                                                 
\end{document}